\documentclass[aps,prb,preprint,superscriptaddress,floatfix,longbibliography,citeautoscript]{revtex4-2}

\usepackage{amsmath,amssymb} 
\usepackage{bm} 
\usepackage{graphicx} 
\usepackage{comment} 
\usepackage[normalem]{ulem} 
\usepackage{setspace}

\usepackage{enumitem}
\setlist{noitemsep,leftmargin=*,topsep=0pt,parsep=0pt}

\usepackage{xcolor} 
\definecolor{lightgray}{gray}{0.6}
\definecolor{medgray}{gray}{0.4}
\definecolor{mRed}{RGB}{230, 0, 50}
\colorlet{newtextColor}{mRed}

\usepackage{hyperref}
\hypersetup{
colorlinks=true,
urlcolor= blue,
citecolor=blue,
linkcolor= blue,
}

\newif\ifptitle
\newif\ifpnumber
\newcounter{para}

\newif\iftrackchanges

\usepackage{mdframed}
\newmdenv[
  linecolor={\iftrackchanges newtextColor\else white\fi},
  linewidth=2pt,
  topline=false,
  bottomline=false,
  rightline=false,
  skipabove=\topsep,
  skipbelow=\topsep,
  leftmargin=-12pt,
  innertopmargin=0pt,
  innerbottommargin=0pt
]{newtextblock}

\ptitletrue  
\pnumbertrue 

\newcommand{\LNO}{La$_{3}$Ni$_{2}$O$_7$\,\,}
\newcommand{\LNOtwo}{La$_{2}$Ni$ $O$_4$\,\,}
\newcommand{\Tc}{$T_\textrm{c}$\,\,}
\newcommand{\LNOfour}{La$_{4}$Ni$_{3}$O$_{10}$\,}

\newcommand{\aep}{School of Applied \& Engineering Physics, Cornell University, Ithaca, NY 14853, USA}

\newcommand{\MSE}{Department of Materials Science and Engineering, Cornell University, Ithaca, NY 14853, USA}

\newcommand{\kavli}{Kavli Institute at Cornell for Nanoscale Science, Cornell University, Ithaca, NY 14853, USA}

\newcommand{\mpi}{Max Planck Institute for Chemical Physics of Solids, 01187 Dresden, Germany}

\newcommand{\harvardphys}{Department of Physics, Harvard University, Cambridge, MA 02138, USA}

\newcommand{\harvardeng}{John A. Paulson School of Engineering and Applied Sciences, Harvard University, Cambridge, MA 02138, USA}

\newcommand{\slac}{Stanford Institute for Materials and Energy Sciences, SLAC National Accelerator Laboratory, Menlo Park, CA 94025, USA}

\newcommand{\stanford}{Department of Applied Physics, Stanford University, Stanford, CA 94305, USA}

\newcommand{\stanfordPhys}{Department of Physics, Stanford University, Stanford, CA 94305, USA}

\newcommand{\asu}{Department of Physics, Arizona State University, Tempe, AZ 85287, USA}

\newcommand{\mytitle}{Resolving Structural Origins for Superconductivity in Strain-Engineered \LNO Thin Films}

\begin{document}

\title{\mytitle}

\author{Lopa Bhatt}
\affiliation{\aep}

\author{Abigail Y. Jiang}
\affiliation{\harvardphys}
\affiliation{\harvardeng}

\author{Eun Kyo Ko}
\affiliation{\slac}
\affiliation{\stanford}

\author{Noah Schnitzer}
\affiliation{\MSE}
\affiliation{\kavli}

\author{Grace A. Pan}
\affiliation{\harvardphys}

\author{Dan Ferenc Segedin}
\affiliation{\harvardphys}

\author{Yidi Liu}
\affiliation{\slac}
\affiliation{\stanfordPhys}

\author{Yijun Yu}
\affiliation{\slac}
\affiliation{\stanford}

\author{Yi-Feng Zhao}
\affiliation{\asu}

\author{Edgar Abarca Morales}
\affiliation{\mpi}

\author{Charles M. Brooks}
\affiliation{\harvardphys}

\author{Antia S. Botana}
\affiliation{\asu}

\author{Harold Y. Hwang}
\affiliation{\slac}
\affiliation{\stanford}

\author{Julia A. Mundy}
\affiliation{\harvardphys}
\affiliation{\harvardeng}

\author{David A. Muller}
\affiliation{\aep}
\affiliation{\kavli}

\author{Berit H. Goodge}
\email{berit.goodge@cpfs.mpg.de}
\affiliation{\mpi}

\date{\today}

\begin{abstract}

The discovery of high-temperature superconductivity in bulk \LNO under high hydrostatic pressure and, more recently, biaxial compression in epitaxial thin films has ignited significant interest in understanding the interplay between atomic and electronic structure in these compounds.  
Subtle changes in the nickel-oxygen bonding environment are thought to be key drivers for stabilizing superconductivity, but specific details of which bonds and which modifications are most relevant remains so far unresolved.
While direct, atomic-scale structural characterization under hydrostatic pressure is beyond current experimental capabilities, static stabilization of strained \LNO films provides a platform well-suited to investigation with new picometer-resolution electron microscopy methods. 
Here, we use multislice electron ptychography to directly measure the atomic-scale structural evolution of \LNO thin films across a wide range of biaxial strains tuned via substrate. 
By resolving both the cation and oxygen sublattices, we study the strain-dependent evolution of atomic bonds, providing the opportunity to isolate and disentangle the effects of specific structural motifs for stabilizing superconductivity.
We identify the lifting of crystalline symmetry through modification of nickel-oxygen octahedral distortions under compressive strain as a key structural ingredient for superconductivity. 
Rather than previously supposed $c$-axis compression, our results highlight the importance of in-plane biaxial compression in superconducting thin films, which suggests an alternative -- possibly cuprate-like -- understanding of the electronic structure.
Identifying local regions of inhomogeneous oxygen stoichiometry and high internal strain near crystalline defects, we suggest potential pathways for improving the sharpness and temperature of the superconducting transition.

\end{abstract}

\maketitle

\newpage
\subsection*{Introduction}

Attempts to identify additional high-temperature superconductors have occupied significant experimental and theoretical efforts since the discovery of superconductivity in cuprates.
The nickelate materials have been a sustained target, and superconductivity was discovered in thin films of the square-planar nickelates after a many-decade search \cite{Li2019}.  
Subsequently, high-temperature superconductivity was also observed in the layered $n =2, 3$ Ruddlesden-Popper nickelates, La$_{n+1}$Ni$_n$O$_{3n+1}$, under high hydrostatic pressure \cite{sun2023signatures,zhu2024superconductivity,wang2024bulk}.  
With maximum superconducting transition temperatures (\Tc) surpassing 80 K in bilayer ($n =2$) compounds near 20 GPa \cite{sun2023signatures, Zhang2024,wang2024bulk}, the nickelates are clearly a true family of high-temperature superconductors.  
Still, the high pressures required to achieve the superconducting state limit both the accessible experimental probes -- especially those of, e.g., electronic band structure and microscopic properties -- and potential applications of the Ruddlesden-Popper nickelates. 
Most recently, epitaxial stabilization of compressively strained \LNO thin films revealed a superconducting transition onset above 40 K \cite{ko2024signatures}, providing a promising pathway for both studying and exploiting high-temperature nickelate superconductivity at ambient pressure \cite{ko2024signatures, zhou2024ambient}.

The square-planar $R$NiO$_2$ ($R$ = hole-doped rare earth) \cite{wang2024experimental} and related Nd$_6$Ni$_5$O$_{12}$ \cite{pan2022superconductivity} nickelate compounds have a nickel $d^{9-\delta}$ electron filling, isoelectronic to that of the cuprates.  
Theoretical work and experimental measurements have suggested a predominant nickel $d_{x^2-y^2}$ band at the Fermi level and a superconducting state with $d$-electron pairing \cite{ wu2020robust, pan2022superconductivity, Ding2024, sun2024electronic, cheng2024evidence}.  
Although there remain some key distinctions \cite{GoodgeOKPNAS} and outstanding theoretical questions, these materials appear to be in many ways very cuprate-like \cite{Lee2023}.  
In contrast, links to the Ruddlesden-Popper nickelates of bilayer La$_3$Ni$_2$O$_7$ (Fig. 1a, $d^{7.5}$) and trilayer La$_4$Ni$_3$O$_{10}$ ($d^{7.33}$) that superconduct under pressure are less obvious: the proposed Fermi surface includes prominent contributions from both the nickel $d_{z^2}$ and $d_{x^2-y^2}$ orbitals \cite{luo2023bilayer, zhang2023electronic, sakakibara2024, christiansson2023correlated, lechermann2023electronic, Lu2024, yang2023possible}.   
It has been hypothesized that the relevant role of different orbitals may lead to competition between different
pairing symmetries from $d$-wave to $s^{\pm}$ \cite{liu2023role, zhang2024structural}.
The emergence of superconductivity at low temperatures may be also linked to the concommitant suppression of a density-wave state \cite{Meng2024}.

So far, the relative roles of structural and electronic reconfiguration--and their interplay--for driving superconductivity in \LNO under pressure remain unclear.
The recent observation of superconductivity in compressively strained La$_3$Ni$_2$O$_7$ thin films thus provides a powerful platform to isolate and thereby disentangle specific structural or electronic contributions for superconductivity in these materials. 
For example, some theoretical calculations focused on the structural symmetry change under pressure have suggested uniaxial or biaxial compression as a promising route to modify the Ni-O$_6$ octahedral patterns \cite{rhodes2024structural}.
Other proposals predicted that superconductivity might emerge under biaxial tensile strain in thin films as the corresponding compression of the $c$-axis would give rise to a Fermi surface which mirrors what is predicted under hydrostatic pressure \cite{geisler2024}. 
Further work highlighted the potential to decouple structural and electronic ingredients necessary for superconductivity between compressive and tensile strain limits \cite{zhao2024electronic}, and proposed strain as a route to tune the superconducting pairing symmetry \cite{zhao2024electronic, luo2024, liu2023role}. 
In this framework, epitaxial thin films provide a uniquely suited platform to study statically stabilized structures as a function of biaxial pressure which are accessible for measurements with the highest real-space resolution.

Here, we leverage scanning transmission electron microscopy (STEM) to provide key insights into the atomic-scale structure of \LNO thin films with varying epitaxial strain. 
Using high-resolution STEM imaging and the newly developed technique of multislice electron ptychography (MEP) \cite{ZhenMEPScience,zhen327Nature}, we directly characterize a range of structural order parameters including Ni-O bond angles and lattice constants across a complete strain series to isolate the structural motifs common in superconducting \LNO. 
Our measurements reveal that under compressive (tensile) strain, the Ni-planar O bond coordination resembles those reported from x-ray refinements in the high-pressure (low-pressure) bulk 
structures (Fig. 1b) -- here imaged directly in real-space for the first time -- 
pointing to the importance of raising crystalline symmetry for superconductivity in these materials. 
Quantitative parametrization of lattice constants also reveals the importance of in-plane rather than out-of-plane compression as has been previously proposed in the context of Ni $d_{z^2}$-driven superconductivity.
We also identify local regions of high internal crystalline strain and stoichiometric inhomogeneity which may be linked to the broad superconducting transition so far observed within thin film samples \cite{ko2024signatures, zhou2024ambient}, suggesting promising routes for enhancing superconductivity in the future. 
This detailed structural study of \LNO thin films across a range of compressive and tensile strain sheds light on the complex relation between structural and electronic degrees of freedom in this new class of high-temperature superconductors.

\subsection *{ Electrical properties of strain-engineered thin films}

\begin{figure*}[t]
    \includegraphics[clip=true,width=\columnwidth]{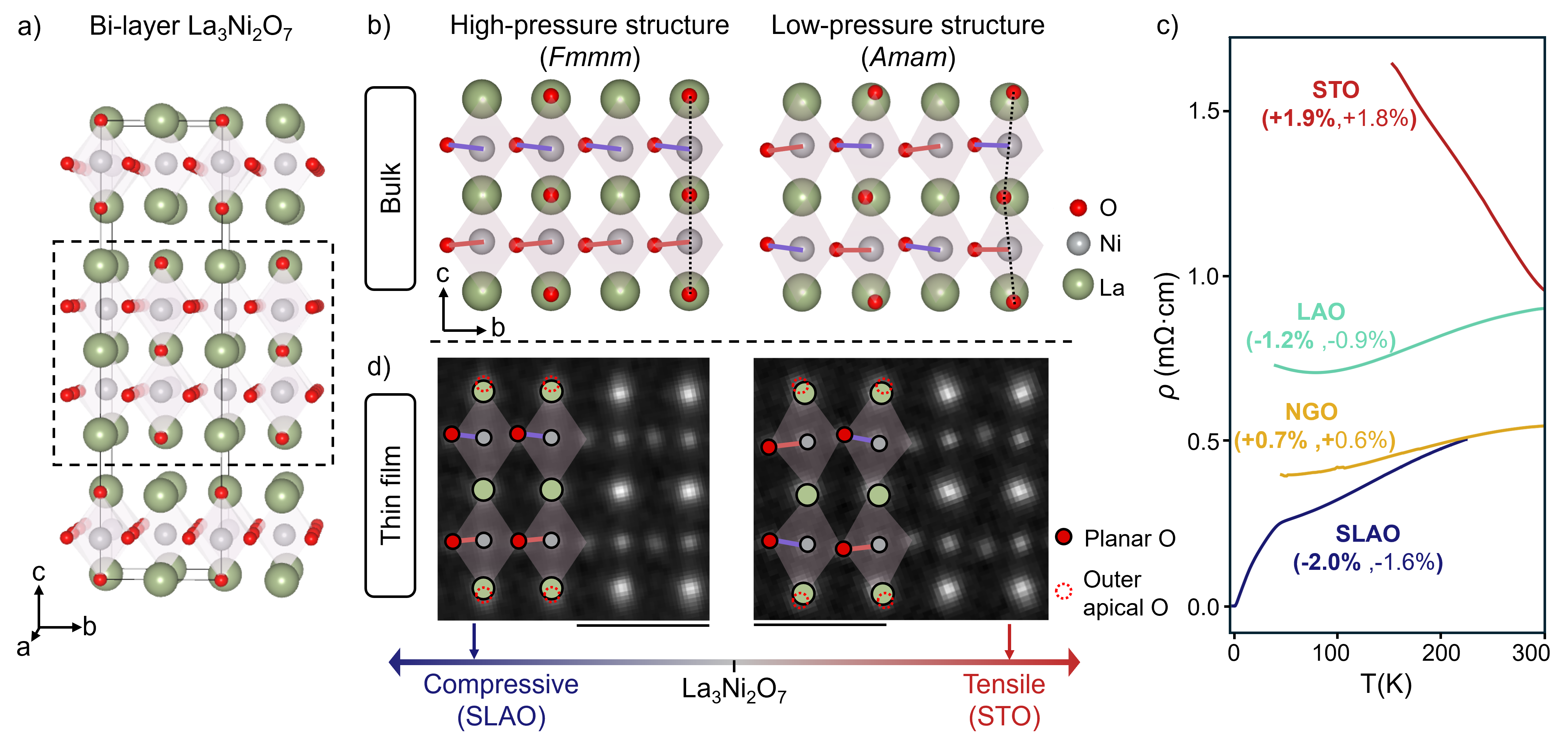}
    \caption{\textbf{Structural and electrical changes in \LNO thin films under compressive and tensile strain.} \textbf{a)} Schematic model of high-pressure $Fmmm$ \LNO with the unit cell marked by a grey box. The dashed box outlines a single bilayer of \LNO. \textbf{b)}  High-pressure and low-pressure bulk structures of one bilayer reported by Sun, et al. \cite{sun2023signatures}. Dashed lines mark the Ni-apical O bond angles; Ni-planar O bond angles are highlighted by orange and purple lines.  
    \textbf{c)} Resistivity $\rho$(T) of \LNO thin films epitaxially strained on substrates SLAO (blue), LAO (cyan), NGO (yellow), and STO (red); chemical formulas are given in the text. The nominal (bold) and measured epitaxial strains imparted on each film are given in parentheses. The film grown on SLAO shows a superconducting onset near 42 K \cite{ko2024signatures}.
    \textbf{d)} Atomic structures of \LNO thin films under compressive (left) and tensile (right) strain measured experimentally by multislice electron ptychography with colored circles identifying the La (green), Ni (grey), planar O (filled red), and outer apical O (open dashed red) atomic columns. Scale bars are 5 \AA.
     \label{fig:strain}}
\end{figure*}

Epitaxially strained thin films of La$_3$Ni$_2$O$_{7-\delta}$ are synthesized on substrates SrLaAlO$_4$ (SLAO), LaAlO$_3$ (LAO), NdGaO$_3$ (NGO), and SrTiO$_3$ (STO) with pseudo-cubic in-plane lattice constants of 3.756 \AA, 3.787 \AA, 3.858 \AA, and 3.905 \AA, respectively (Methods). 
All films are treated with post-growth annealing in molecular oxygen or ozone with the goal of obtaining oxygen occupancy close to stoichiometric \LNO, though we note that the precise quantification of oxygen stoichiometry in both thin films and bulk crystals is experimentally challenging \cite{zhen327Nature}. 
Following $\varepsilon = (a_{sub}-a_{bulk})/a_{bulk}$, the changes of the in-plane lattice constant compared to ambient-temperature, ambient-pressure bulk \LNO (pseudo-tetragonal $a_{pt} =  \sqrt{a^2  + b^2}/2$ = 3.833 \AA \, \cite{ko2024signatures}) impart nominal epitaxial strains of -2\% (SLAO), -1.2\% (LAO), +0.7\% (NGO) and +1.9\% (STO), though the experimentally measured values which show some lattice relaxation in films are given in Fig. 1c and Supplemental Table 1.
Here, negative (positive) values denote compressive (tensile) strain. 
For consistency, we use the lattice convention shown in Fig. 1a where $c$ is the long axis along the layer stacking direction and refer to nominal strain values for simplicity.
Films grown on SLAO, LAO, and NGO yield high-quality crystalline Ruddlesden-Popper nickelates with intermittent mixed-phase intergrowth layers, mainly of \LNOtwo, and extended defects spanning the films (Supplemental Figs. S1-3), while films grown on STO show the highest density of extended defects due to the high tensile strain imposed by the substrate \cite{ferenc2023limits, Cui2024StrainMediated}.

Electronically, the film on SLAO under highest compressive strain shows a broad superconducting transition at ambient pressure with onset close to 42 K, reaching zero resistance around 2 K (Fig. 1c) as described previously \cite{ko2024signatures}. 
Films grown on LAO and NGO show metallic behavior while films on STO are insulating (Fig. 1c), consistent with previous studies and across different growth methods \cite{ko2024signatures,Cui2024StrainMediated} (Supplemental Fig. S4).
This clear change in the electronic properties from insulating to superconducting with increasing compressive strain demonstrates the impressive tunability of complex nickel oxides.

\subsection*{Strain-dependent Ni-O bond symmetry}

The compression-induced straightening of the Ni-apical O bond angle in bulk \LNO (Fig. 1b) thought to modify the Ni-O hybridization and Fermi surface structure correlates to a symmetry raising from orthorhombic $Amam$ to $Fmmm$ or tetragonal $I4/mmm$ \cite{sun2023signatures, wang2024structure, li2024pressure}. 
Indeed, the variability of crystalline symmetries reported highlights the experimental difficulty to distinguish similar phases especially in the presence of mixed structural domains or twins \cite{li2024pressure, wang2024structure, zhou2024revealing, wang2024bulk}, and emphasizes the need for direct, real-space measurements of the octahedral structure (Supplemental Fig. 5). 
Quantitative measurements of light element positions -- especially in the presence of heavy ions -- with high spatial resolution and precision is, however, experimentally challenging. 
Local cross-sectional imaging in the scanning transmission electron microscope (STEM) provides a platform for directly measuring atomic structure in thin films, but particular care must be taken for robust measurements of light elements such as oxygen. 
While certain collection geometries such as annular bright-field (ABF) or integrated differential phase contrast (iDPC) can resolve light element positions in carefully prepared specimens, they are susceptible to artifacts based on the experimental set-up (e.g., crystalline projection alignment, aberrations in the imaging probe) and sample geometry (e.g., specimen thickness, electron dechanneling near defects) which hinder quantitative measurements \cite{Guo2018ABFArtifact,Burger2020DPC}.

Multislice electron ptychography (MEP) in the STEM is a recently implemented imaging technique that allows for quantitative measurement of light atoms while eliminating many such artifacts. 
By solving the inverse problem of electron scattering to attain the sample potential, MEP provides deep sub-\AA ngstr\"om spatial resolution, depth-resolved information along the projection direction, and interpretable contrast spanning both light and heavy atoms \cite{ZhenMEPScience}.
As in conventional high-resolution STEM imaging, a focused probe of high-energy electrons is rastered across the sample (Supplemental Fig. S6).
Compared to conventional imaging modes which record only a single intensity at each probe position, a full diffraction pattern is recorded at every real-space position using a high-dynamic range pixel array detector \cite{Yi20182DPtycho}. 
The phase information of the scattered electrons encoded in the overlapping diffraction disks is used to reconstruct the sample potential through iterative algorithms \cite{Yi20182DPtycho,ZhenMEPScience}.
MEP therefore offers a unique capability to quantify subtle structural distortions in the Ni-O$_6$ octahedra across the series of strained \LNO thin films.

\begin{figure*}[t]
    \includegraphics[clip=true,width=\columnwidth]{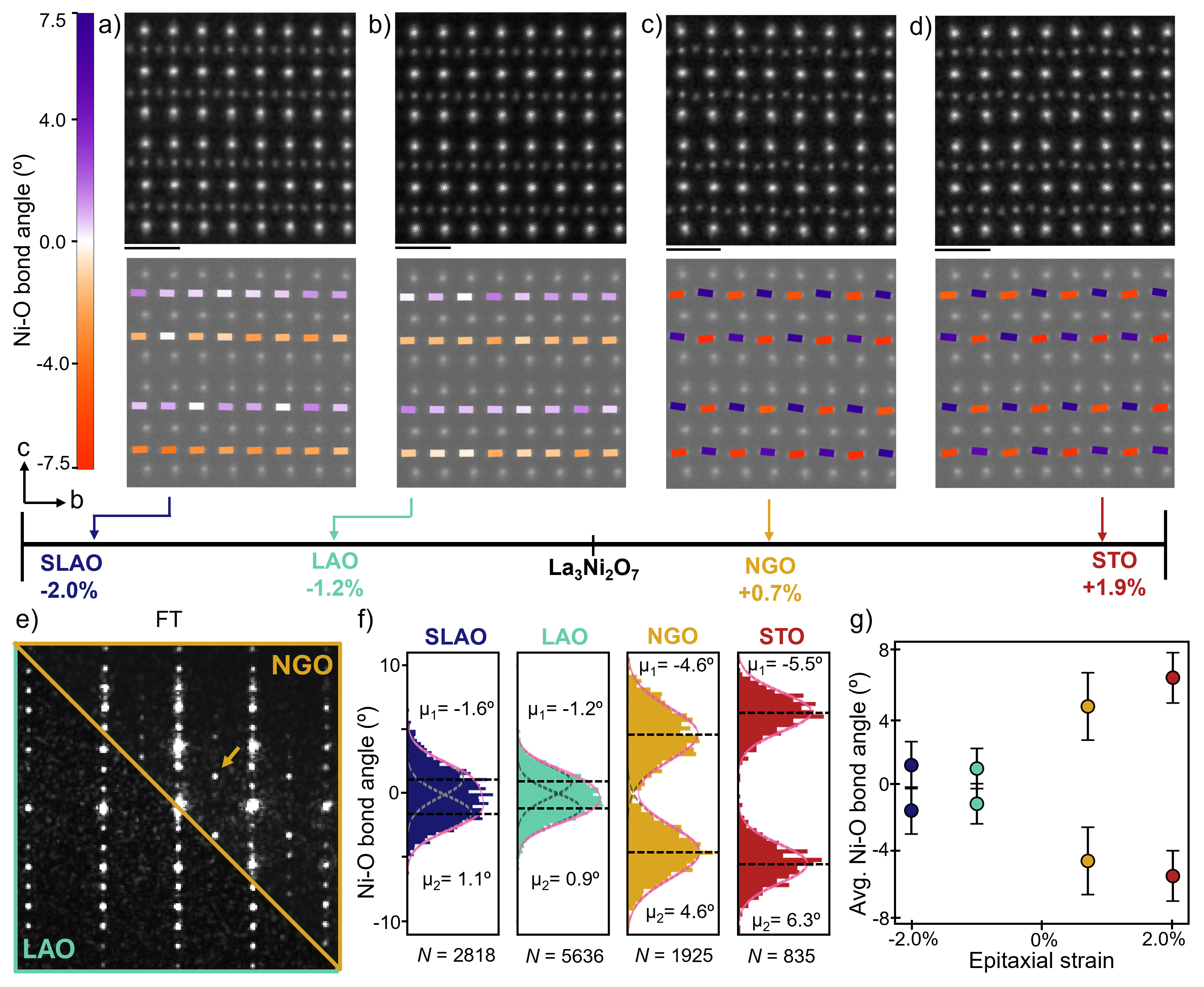}
    \caption{\textbf{Ni-O bond angle evolution across epitaxial strain in \LNO thin films.}
    \textbf{a-d)} MEP reconstructions (top) and Ni-planar O bond angle maps (bottom) of \LNO thin films grown on SLAO, LAO, NGO, and STO substrates. \textbf{e)} Fourier transforms (FT) of ADF images of thin films on NGO (upper right) and LAO (bottom left). Differences in symmetry between the tensile and compressively strained films are apparent by additional superlattice peaks which indicate a lower crystalline symmetry. 
    \textbf{f)} Ni-planar O bond angles of all four thin films. The gray dashed lines are two Gaussian fits to the bond angle distribution of each film. The pink solid line shows the sum of both Gaussian fits. The mean values ($\mu$) and number of bonds analyzed ($N$) are given for each film. \textbf{g)} Average Ni-planar O bond angle for films under different epitaxial strains. Scale bars are 5 \AA.}
     \label{fig:angles}
\end{figure*}

Figure 1d shows MEP reconstructions of small regions in compressive and tensile strained \LNO thin films projected along the orthorhombic $a$-axis ([100]$_{o}$). 
Lanthanum, nickel, and planar oxygen atomic columns are clearly visible with high spatial resolution. 
Apical O are visible as faint tails of outer La atoms, as indicated by dotted red circles (Fig. 1d). 
Qualitatively, the tails are aligned vertically with Ni atoms in compressive films, while tilted to alternating sides of the La atoms in tensile strained films. 
This suggests that the Ni-apical O bond angles are closer to 180 degrees in the compressive films than in the tensile films. 
A similar trend was identified in density-functional theory-based structural relaxations \cite{rhodes2024structural, zhao2024electronic}.
Robust experimental quantification of the precise vertical bond angles is, however, problematic due to the projection nature of STEM imaging: although signatures of the apical O columns are clearly discernible in the orthorhombic projection shown here, their overlap with La columns makes precise measurement of the atomic positions challenging. 
The O and La columns are easily distinguished along the [110]$_{o}$ projection, but in this orientation the octahedral rotations in the $Amam$ space group are antiphase stacked which results in an apparent ellipticity of O sites in incoherently distorted octahedra (Supplemental Fig. S6) \cite{zhen327Nature}.

To overcome this challenge, we focus instead on the Ni-planar O bond angles which also rearrange in response to pressure as a way to fingerprint structural symmetry changes in thin films across varying strain.
We quantify local bond angles by first obtaining Ni and planar O positions through a 7-parameter Gaussian fitting of the atomic columns in MEP reconstructions, which are then used to calculate the projected Ni-O bond angle with respect to the plane between neighboring Ni columns (Supplemental Fig. S5,7).

Figure 2a-d (top) show MEP reconstructions of \LNO thin films spanning epitaxial strain from -2\% to +2\%, here each cropped to two bilayers for display purposes.
Maps of the individual bond angles overlaid on the MEP reconstructions are presented in the bottom row. 
Most striking is the qualitative difference between the symmetry of compressive and tensile strained films: across a single Ni-O$_2$ plane in each bilayer, the Ni-planar O bonds are aligned unidirectionally in the compressive films but alternate in tensile films. 
This change in symmetry is also visible through the presence of extra peaks in Fourier transforms of annular dark-field (ADF)-STEM images of each sample (Fig. 2e, Supplemental Fig. S8). 
The additional half-order peaks visible in tensile strained films (yellow arrow in Fig. 2e) and lack thereof in compressive films indicate relatively lower and higher crystalline symmetries respectively, and are in good agreement with the expected patterns for structures reported for \LNO under low-pressure and high-pressure conditions \cite{sun2023signatures,wang2024structure}.

Here, the observed octahedral distortions in the compressively strained films shows a close match to the originally reported $Fmmm$ structure \cite{sun2023signatures} in which the planar-O shift away from the center La-O plane (Supplemental Fig. S7).
Theoretical calculations also predict compressive stabilization of $Fmmm$ symmetry (Supplemental Fig. S9) \cite{rhodes2024structural}. 
We note that assuming equivalent $a$ and $b$ lattice constants imposed by the SLAO substrate lifts orthorhombic $Fmmm$ to a tetragonal space group consistent with $I4/mmm$ \cite{wang2024bulk}. 
Still, we emphasize that the Ni-O bond distortions observed here are distinct from those in reported bulk $I4/mmm$ refinements (Supplemental Fig. S7) \cite{wang2024structure}.
Temperature-dependent experiments have suggested an additional structural transition under high pressure conditions \cite{wang2024structure}, which in atomic models correspond to an inversion of the Ni-planar O bond orientation (Supplemental Fig. S7).
Whether the superconducting thin film undergoes a similar low-temperature structural transition remains an open question, beyond the current limitations of cryogenic electron microscopy \cite{bianco2021atomic}.

In addition to this change in symmetry, MEP measurements also reveal a systematic trend in the magnitude of the bond angles across the strain series.
The magnitude average bond angle measured from several MEP reconstructions spanning larger total areas of each film increases with increasing tensile strain (Fig. 2f,g, Supplemental Fig. S10). 
Interestingly, despite their similar symmetry, the bond angles measured in the superconducting film on SLAO are smaller than those in the reported high-pressure refinements \cite{sun2023signatures,wang2024structure} (Supplemental Fig. S7), a trend which is reproduced with calculated relaxations of $Fmmm$ \LNO biaxially strained to SLAO and LAO lattice constants (Supplemental Fig. S11, Supplemental Table 2,3).

\subsection*{Lattice evolution across the strain series}

In addition to understanding subtle changes in the symmetry of Ni-O bonds upon compression, we also analyze the in-plane (IP) and out-of-plane (OOP) lattice constants probed both globally by x-ray diffraction (XRD) and locally at the atomic scale by ADF-STEM imaging. 
As shown in Fig. 3a, the IP lattice constants measured via STEM and XRD decrease as expected with increasing compressive strain, similar to the decreasing trend measured in bulk crystals under hydrostatic pressure (Fig. 3d). 
The IP spacing of the superconducting film on SLAO closely matches the IP spacing of bulk samples at the superconducting critical pressure. 
Notably, the trend inverts for the OOP lattice constant in thin films (Fig. 3b, Supplemental Fig. S12), where the Poisson effect drives an expansion (compression) of the unconstrained lattice dimension under biaxial compression (tension). 
In particular, the $c$-axis spacing increases under compression such that the superconducting thin film on SLAO has an OOP lattice spacing furthest from that of bulk samples which superconduct under high pressure (Fig. 3e). 
The unit cell volume of the strained films (Fig. 3c) decreases with compression, though even at the highest compression in the superconducting film on SLAO the volume is significantly larger than superconducting bulk samples under hydrostatic pressure (Fig. 3f), also in agreement with theoretical relaxations (Supplemental Tables 1-3).

Taken together, the combined parametrization of lattice structures in strained films and pressurized bulk crystals indicates the relevance of the in-plane lattice constant and octahedral symmetry as order parameters for superconductivity. 
To date, a large emphasis has been given to compression of the $c$-axis in relation to superconductivity under pressure and the related emergence of an additional Ni-$d_{z^2}$ state near the Fermi level \cite{sun2023signatures,yang2023possible, zhang2023electronic,Lu2024}. 
Our results clearly show that superconductivity is not driven by $c$-axis compression alone, calling for an alternative electronic understanding of superconductivity in these materials. 
In this framework, recent work has shown that compressive strain, on the other hand, might suppress the additional Ni-$d_{z^2}$ state from the Fermi surface and lead to a more cuprate-like electronic structure \cite{zhao2024electronic, geisler2024}.

\begin{figure*}[t]
    \includegraphics[clip=true,width=\columnwidth]{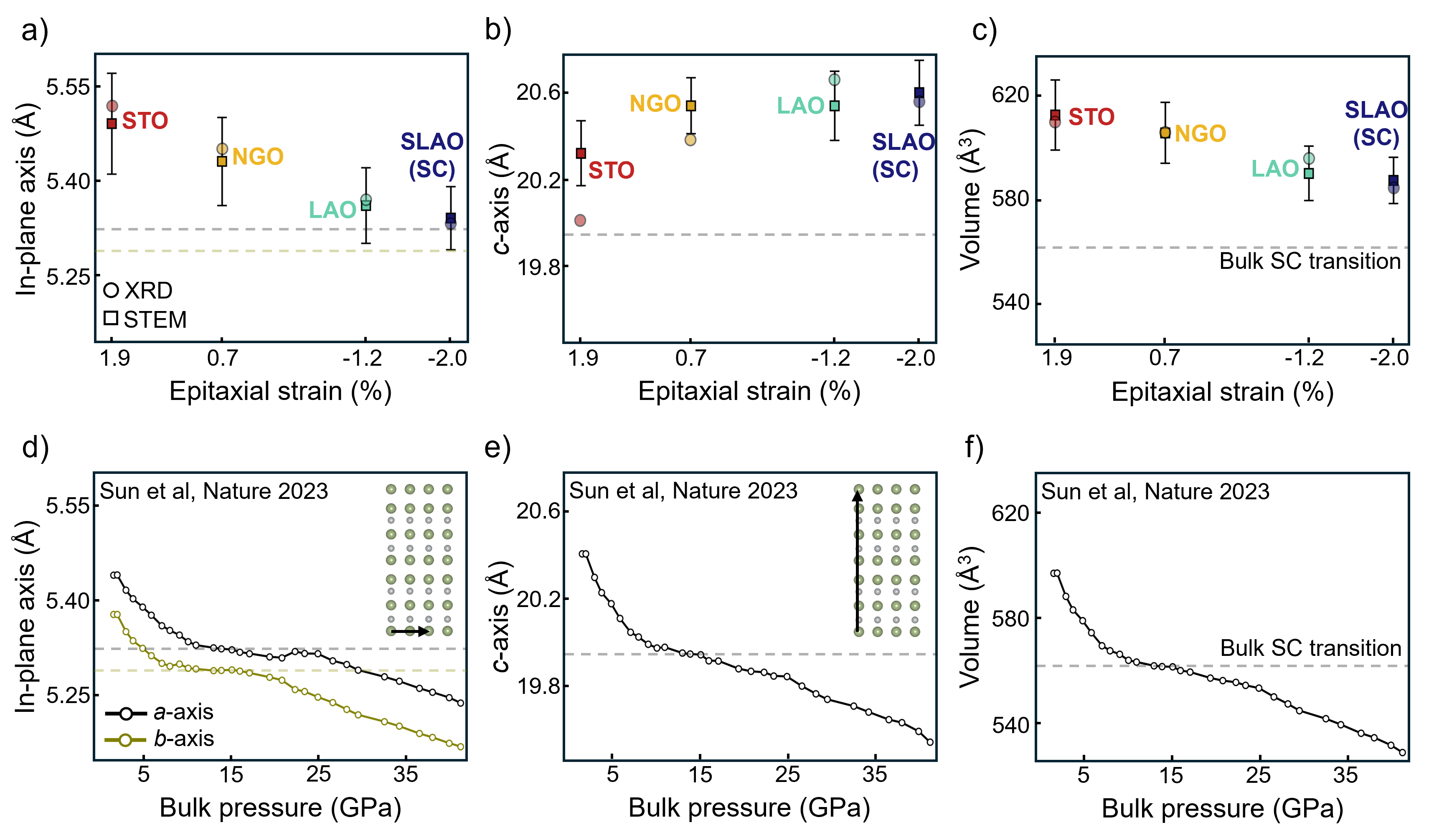}
    \caption{\textbf{Lattice parameter changes across epitaxial strain in \LNO thin films.}
     \textbf{a-c)} In-plane lattice constant (a), $c$-axis lattice constant (b), and unit cell volume (c) of thin films as a function of epitaxial strain measured via XRD (circles) and ADF-STEM (squares). Dashed lines indicate the reported values of each parameter at the critical pressure for superconductivity.
     \textbf{d-f)} In-plane lattice constant (d), $c$-axis lattice constant (e), and unit cell volume (f) versus pressure in bulk \LNO reproduced from Sun, et al. \cite{sun2023signatures}.  
}
     \label{fig:params}
\end{figure*}

\subsection*{Secondary phases and their impact on superconductivity}

\begin{figure*}[t]
    \includegraphics[clip=true,width=6 in]{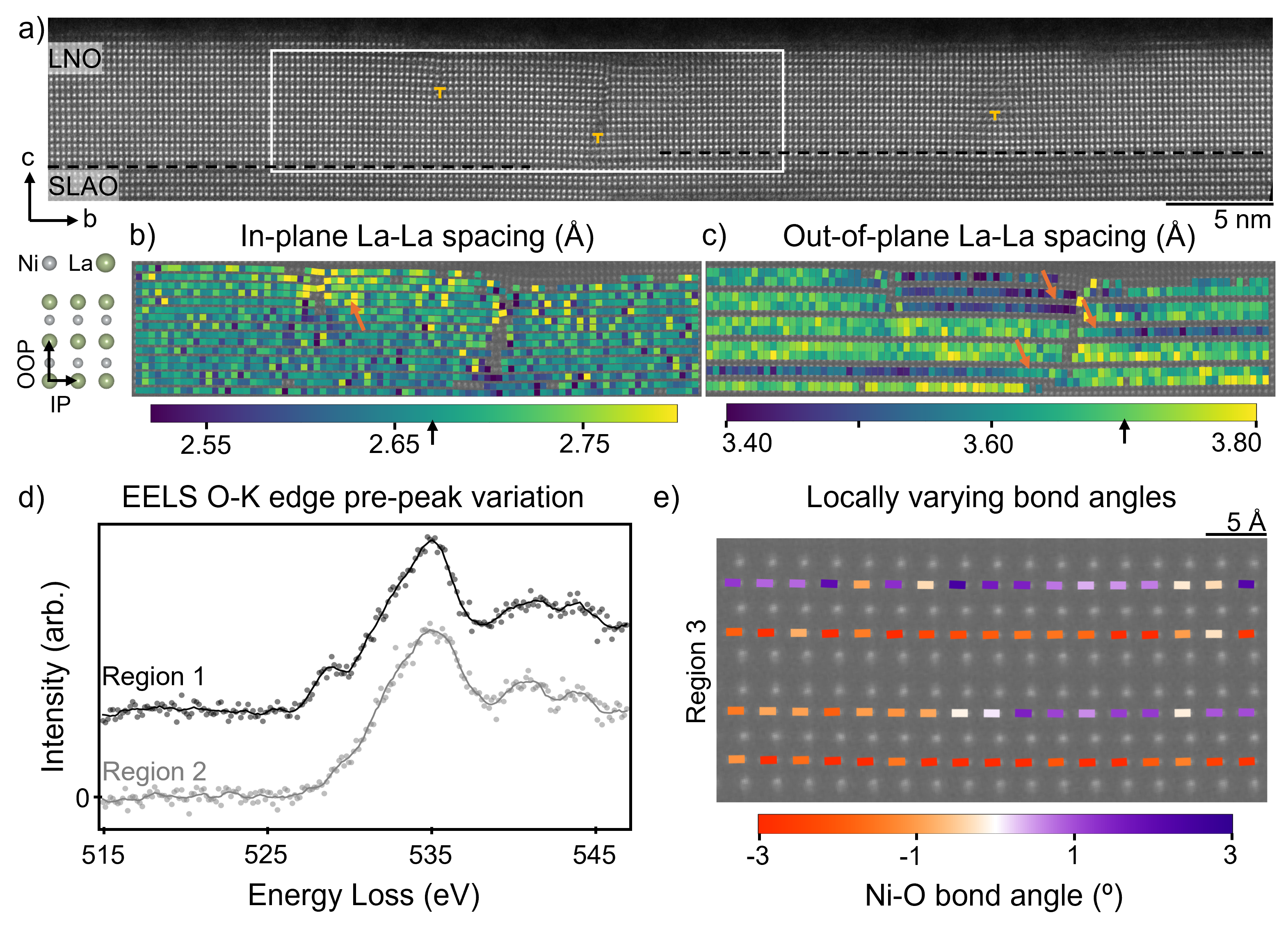}
    \caption{ \textbf{Impact of local inhomogeneities on lattice and oxygen structure.} \textbf{a)} Large field-of-view ADF-STEM image of the \LNO film on SLAO. Dashed lines mark the interface between film and substrate. Yellow $\top$s mark crystalline dislocations. \textbf{b-c)} In-plane (b) and out-of-plane La-La spacing (c) measured by quantitative atom-tracking in the region marked by white box in (a). Orange arrows mark relaxation near defects. Average IP and OOP La-La spacings in clean regions of the film are marked by black arrows on the colorbar. \textbf{d-e)} O-K edge ELNES (d) and Ni-planar O bond angle map (e) from three different regions of the same \LNO film on SLAO (ADF images of regions 1 and 2 are shown in Supplemental Fig. S15).}
     \label{fig:defects}
\end{figure*}

A key distinction between the superconductivity reported in thin film and bulk \LNO is the breadth of the superconducting transition, which is significantly wider in films than in bulk \cite{ko2024signatures,zhou2024ambient}. 
It should be noted that comparison of in-plane lattice constants maps the biaxial strain on SLAO to approximately 10-20 GPa biaxial pressure, near the edge of the superconducting boundary in the pressure-temperature phase diagram of bulk crystals where the superconducting transition is also broad \cite{GWangPRX2024,sun2023signatures,Zhang2024}. 
Indeed, the critical compressive strain required to stabilize superconductivity in \LNO thin films has yet to be systematically established: future synthesis efforts to traverse a range of compressive strains in high-quality thin films and theoretical modeling will likely play key roles in this regard.

While the lattice parameters in Fig. 3 represent an ensemble understanding of the film structure, epitaxial relaxation and crystalline defects can locally modify these parameters such that they may fall outside the window required for superconductivity. 
Figure 4a shows a large field-of-view ADF-STEM image of the superconducting \LNO thin film on SLAO. 
Near strain-relieving dislocations marked by yellow $\top$, the IP La-La distances measured by quantitative atom-tracking (Fig. 4b) show clear expansion away from the average compressed value of $\sim$2.67 \AA\, marked by an arrow on the color bar. 
Other signs of minor relaxation can also be observed by the corresponding reduction of the OOP La-La distances near the surface of the film (Fig. 4c).
A second crystalline defect which extends throughout the vertical direction of the film can be traced to a step-edge in the SLAO substrate (Supplemental Fig. S13), and correlates with stacking variation on either side. 
Single layers of the $n=1$ Ruddlesden-Popper phase \LNOtwo are especially visible by their reduced OOP La-La spacing (Fig. 4c).
Similar phase variations and their links to substrate surface variation have been documented in other Ruddlesden-Popper thin films \cite{kim2021superconducting}. 
Tensile-strained films also show in-plane crystalline twin domains where the inequivalent $a$- and $b$-axes can be distinguished by their distinct octahedral distortions (Supplemental Fig. S14).

Intergrowths of $n=1$ and other Ruddlesden-Popper phases are observed in all films within our series and were included in previous reports of superconducting \LNO thin films \cite{ko2024signatures}. 
Similar intergrowths have also been documented in bulk crystals of \LNO \cite{puphal2024unconventional,chen2024polymorphism, wang2024long, li2024design}, though the importance for superconductivity of such polymorphs along the vertical stacking direction has not been established.  
When different intergrowths meet laterally in the film, however, the large internal crystalline strain can lead to significant local modification of the lattice constants, as visible in several regions of Fig. 4b,c.
We speculate that these regions of concentrated internal strain may have a stronger impact on the macroscopic properties of the very thin films studied here, while similar effects may be more dilute in bulk sample volumes.
These observations also provide a tantalizing hint towards the possibility of engineering filamentary superconductivity at highly strained heterostructure interfaces or grain boundaries in bulk crystals. 

In addition to crystalline defects and stacking variation, more subtle point-like disorder such as atomic vacancies may also have a strong impact on superconductivity. 
Previous studies in both bulk and thin film \LNO established the importance of oxygen stoichiometry: bulk samples which are oxygen-deficient exhibit insulating behavior \cite{zhang1994synthesis, taniguchi1995transport,  kobayashi1996transport}, and thin films on SLAO require additional post-growth ozone annealing to exhibit a superconducting transition \cite{ko2024signatures, zhou2024ambient}.

We investigate the oxygen occupancy at mesoscopic length scales by electron energy loss spectroscopy (EELS) which provides access to the local chemical and electronic structure \cite{kourkoutis2010atomic}.
The O-K edge energy loss near edge structure (ELNES) structure fingerprints the hybridization between Ni-3$d$ and O-2$p$ states and has proven a reliable method to measure oxygen stoichiometry in nickelates \cite{zhen327Nature, GoodgeOKPNAS}, particularly by tracking the strength of the pre-peak spectral feature. 
Figure 4d shows two EEL spectra collected on the same day from two highly-crystalline regions of the same \LNO thin film on SLAO (Supplemental Fig. S15).
One region (black) shows a pronounced pre-peak of the O-K edge around 528 eV consistent with previous measurements of near-stoichiometric \LNO \cite{zhen327Nature}, while in a second region (gray) the pre-peak is absent suggesting a reduced oxygen occupation \cite{zhen327Nature,GoodgeOKPNAS}.

It is likely that concentrated oxygen vacancies may also modify the local octahedral coordination.
Even in pristine bilayers of the superconducting film, we observe regions with local variation of the Ni-planar O bond angle. 
Figure 4e shows one few-nm region of the superconducting film on SLAO where the bond angles follow the unidirectional tilt patterns discussed above on the right side of the field of view, but deviate from this symmetry on the left side, only a few unit cells away. 
These local variations may be due to proximity to defects which result in local strain gradients across several unit cells or the presence of oxygen vacancies which interrupt the octahedral distortion pattern. 
Previous atomic-scale analysis of bulk \LNO showed that oxygen vacancies are concentrated in the middle apical layer \cite{zhen327Nature}, but whether similar accommodation is present in strained thin films and what the effect of such vacancies is on the subtleties of octahedral distortions under different strain states remains an open question. 
Future studies correlating precise structural measurements with oxygen stoichiometry and electronic transport may provide further insights in this regard. 
Together, our observations highlight the tremendous opportunity for improving \LNO thin film quality with the hopes of achieving sharper and higher-temperature superconducting transitions.

Through a systematic study of \LNO thin films grown on different substrates with varying epitaxial strain, we identify essential structural characteristics responsible for electronic changes observed across the strain series. 
Macroscale XRD and quantitative atomic-resolution ADF-STEM measurements point to the necessity of in-plane rather than $c$-axis compression for achieving superconductivity. 
Direct quantitative measurements of Ni-planar O bond angles via MEP reveal a lifting of crystalline symmetry 
under biaxial compressive strain which mirrors that reported in bulk crystals under high pressure conditions. 
Together, these call for revisiting theoretical understanding of the Fermiology in these compounds, suggesting possible cuprate-like physics rather than the prevailing hypotheses of Ni 3$d_{z^2}$ orbital overlap.
We also identify sources of internal strain within the films and key areas for materials improvements which may lead to higher and sharper superconducting transitions. 
Future exploration of the structural changes tuned by even larger compressive strains, chemical pressure, and at low temperatures will further facilitate the quest for understanding structural and electronic interplay in high-temperature superconductors.

\section*{Methods}

\subsection*{Thin film synthesis}

The preparation of \LNO thin films on SLAO(001) substrates via pulsed laser deposition (PLD) are detailed in \cite{ko2024signatures}.
Epitaxial thin films of \LNO on LAO(100), NGO(110), and STO(001) were grown via ozone-assisted molecular beam epitaxy (MBE) following the preparation in \cite{pan2022superconductivity, ferenc2023limits}, with improved structural quality achieved by shuttering LaO and NiO$_2$ monolayers according to \cite{nie2014atomically}. The films were annealed in an OTF-1200X-S vacuum tube furnace (MTI Corporation) under 1 atm O$_2$ pressure for 3 hours at 500 \textdegree C.

\subsection*{Bulk characterization}
The structural, and electronic characterization of \LNO thin films on SLAO(001) substrates are detailed in \cite{ko2024signatures}.
For the thin films of \LNO on LAO(100), NGO(110), and STO(001), XRD was collected on a Malvern Panalytical Empyrean diffractometer with Cu K$\alpha_1$ radiation ($\lambda$ = 1.5406 \AA). $c$-axis lattice parameters were calculated with Nelson-Riley fitting \cite{nelson1945fit}, and in-plane parameters were obtained from reciprocal space maps collected with a PIXcel3D area detector. Electrical transport data was collected on a Quantum Design Physical Property Measurement System with ultrasonically wire-bonded aluminum-silicon contacts.

\subsection*{Electron microscopy}
Cross-sectional STEM specimens were prepared by standard focused ion beam (FIB) lift-out method on a Thermo Fisher Scientific G4 UX FIB to projection thicknesses of $<$30 nm.  

ADF-STEM images were acquired using a Cs-corrected Thermo Fisher Scientific (TFS) Spectra 300 X-CFEG operating at 300 kV with probe convergence angle of 30 mrad.
Ptychographic 4D-STEM datasets were collected on the same tool equipped with an EMPAD-G2 detector \cite{EMPAD2} using a probe semi-convergence angle of 30 mrad and outer collection angles of $\sim$ 55 or 66 mrad. 
We used step sizes of 0.44 or 0.63 \AA, probe overfocus of 5-15 nm, and a dwell time of 100 ${\mu}s$ per scan position.

The maximum likelihood algorithm implemented in the fold-slice package was used to perform iterative phase retrieval \cite{ZhenMEPScience,Thibault2012Maximumlikelyhood,Wakonig2020Ptychoshelves}, using position correction and multiple probe modes to account for partial coherence of the STEM probe \cite{Thibault2013Mixprobe,zhen2020MixedState}. 
A Bayesian optimization algorithm with data error as the objective function was used to optimize parameters such as defocus, convergence angle, and rotation angle \cite{chenyu2022BO}. 
Final reconstructions were carried out with 8 probe modes and slice thickness of 0.6-0.9 nm.

Electron energy loss spectroscopy (EELS) of the O-K edge was measured on a second Thermo Fisher Scientific Spectra 300 X-CFEG equipped with a Gatan Continuum spectrometer and camera operating at 120 kV accelerating voltage. 
Probe currents were limited to $<$20 pA to reduce possible beam damage \cite{GoodgeOKPNAS, zhen327Nature}. 

\subsection*{First-principles calculations}
We used density-functional theory (DFT) calculations as implemented in the Vienna ab initio
Simulation Package (VASP) to perform structural optimizations in \LNO under different strain levels (applied to both the $Amam$ and $Fmmm$ structures). To that end, we constrained the in-plane lattice constants to the appropriate biaxial strain level $\epsilon=\frac{a^*(b^*)}{a(b)}$ 
and relaxed both the out-of-plane lattice constant and the internal coordinates. The exchange-correlation functional chosen was the Perdew-Burke-Ernzerhof version of the generalizedgradient approximation (GGA-
PBE). An energy cutoff of 500eV and a $k$-mesh of 8$\times$8$\times$3 were
employed for the different strain levels. We fully
optimized the both internal coordinates and the $c$-axis until the
Hellman-Feynman forces were lower than 1 meV/Å.

\section*{Author contributions}
The project was conceived by L.B., B.H.G., and J.A.M.
Thin films on LAO, NGO and STO were synthesized by MBE by A.Y.J., G.A.P., D.F.S., with assistance from C.M.B. and J.A.M.
Thin films on SLAO and LAO were synthesized by PLD by E.K.K., Y.L., and Y.Y. with supervision from H.Y.H.  
Structural relaxations by DFT were performed by E.A.M, Y.-F.Z., and A.S.B. 
Electron microscopy measurements were conducted by L.B. with input from N.S., B.H.G., and D.A.M. 
The manuscript was written by L.B., B.H.G., A.S.B., and J.A.M. with input from all authors. 

\section*{Materials and correspondence}
Correspondence and material requests should be addressed to Berit H. Goodge at berit.goodge@cpfs.mpg.de.

\section*{Competing interests}
The authors declare no competing interests.

\begin{acknowledgments}

L.B. dedicates this paper to Dr. Lena F. Kourkoutis for her continuous guidance, support, and inspiration. 
This work made use of the Cornell Center for Materials Research shared instrumentation facility.
L.B. and D.A.M. acknowledge support by the NSF Platform for the Accelerated Realization, Analysis, and Discovery of Interface Materials (PARADIM) under cooperative agreement No. DMR-2039380. 
The Thermo Fisher Spectra 300 X-CFEG was acquired with support from PARADIM, an NSF-MIP (DMR-2039380), and Cornell University.
A.Y.J., G.A.P., D.F.S., C.M.B. and J.A.M. acknowledge support from US Department of Energy, Office of Basic Energy Sciences, Division of Materials Sciences and Engineering, under award no. DE-SC0021925. A.Y.J., G.A.P., and D.F.S. acknowledge support from the NSF Graduate Research Fellowship. G.A.P. and A.Y.J. were also supported by the Paul and Daisy Soros Fellowship for New Americans and the Ford Foundation. J.A.M. acknowledges support from a Packard Fellowship and a Sloan Fellowship. 
E.K.K, Y.Y, Y.L. and H.Y.H. acknowledge support from the U. S. Department of Energy, Office of Basic Energy Sciences, Division of Materials Sciences and Engineering (Contract No. DE-AC02-76SF00515) and the Gordon and Betty Moore Foundation’s Emergent Phenomena in Quantum Systems Initiative (grant no. GBMF9072, synthesis equipment). Part of the sample fabrication was conducted at the Stanford Nano Shared Facilities (SNSF), supported by the National Science Foundation under grant ECCS-1542152.
Y.-F.Z. acknowledges support from NSF Grant No. DMR-2323971. A.S.B. was supported by the Alfred P. Sloan Foundation (FG-2022-19086).
B.H.G. was supported by the Max Planck Society and the Schmidt Science Fellows in partnership with the Rhodes Trust.

\end{acknowledgments}

\clearpage
\newpage

\bibliography{refs}

\clearpage
\newpage

\section*{Supplemental Information}

\renewcommand{\thefigure}{S\arabic{figure}}
\renewcommand{\thetable}{S\arabic{table}}
\setcounter{figure}{0}

\begin{table}[h!]
  \begin{center}
    \caption{Experimentally measured structural parameters for the series of strained \LNO films via x-ray diffraction (XRD). Experimental in-plane lattice constants and strain values are extracted from fits to x-ray reciprocal space maps (RSM). The $c$-axis spacing is calculated by Nelson-Riley fits to the XRD maps for films on LAO, NGO, and STO. The $c$-axis spacing for the film on SLAO is measured from the XRD (006) Bragg peak. }
    \label{tab:spacing_xray}
    \begin{tabular}{c|c|c|c|c } 
      \hline
       Substrate (nominal $\varepsilon$) & SLAO (-2.0\%) & LAO (-1.2\%) & NGO (+0.7\%)& STO (+1.9\%) \\
      \hline 
      Experimental strain & -1.6\% & -0.9\% & +0.6\% & +1.8\% \\
      In-plane axis (\AA) & 5.33 & 5.37  &  5.45 & 5.52  \\
      $c$ (\AA) & 20.56 & 20.66 $\pm$ 0.13 &  20.39 $\pm$ 0.06 & 20.01 $\pm$ 0.23 \\
      Volume $a \times a \times c$ \,(\AA$^3$)& 584.63  & 595.97 & 605.98 & 609.99 \\
      \hline
    \end{tabular}
  \end{center}
\end{table}

\begin{table}[h!]
  \begin{center}
    \caption{Experimentally measured structural parameters for the series of strained \LNO films. Average lattice spacings and bond angle values are measured via MEP- and ADF-STEM as detailed in the main text, with errors taken as the standard deviation of all measurements.}
    \label{tab:spacing_stem}
    \begin{tabular}{c|c|c|c|c } 
      \hline
       Substrate (nominal $\varepsilon$) & SLAO (-2.0\%) & LAO (-1.2\%) & NGO (+0.7\%)& STO (+1.9\%) \\
      \hline 
      Ni-planar O bond angle ($^{\circ}$)  & 1.7 $\pm$ 1.2 & 1.4 $\pm$ 1.0 & 4.7 $\pm$ 2.0 & 5.9 $\pm$ 1.6 \\
      In-plane axis (\AA) & 5.34 $\pm$ 0.05 & 5.36 $\pm$ 0.06 & 5.43 $\pm$ 0.07 & 5.49 $\pm$ 0.08 \\
      $c$ (\AA) & 20.6 $\pm$ 0.15 & 20.54 $\pm$ 0.16 &  20.54 $\pm$ 0.13 & 20.32 $\pm$ 0.15 \\
      Volume $a \times a \times c$ \,(\AA$^3$)& 587.42 $\pm$ 8.88  & 590.10 $\pm$ 10.41 & 605.62 $\pm$ 11.69 & 612.45 $\pm$ 13.41 \\
      \hline
    \end{tabular}
  \end{center}
\end{table}

\begin{table}[h!]
  \begin{center}
    \caption{DFT calculated lattice spacing and bond angle values.}
    \label{tab:DFT spacing}
    \begin{tabular}{c|c|c|c|c } 
      \hline
      Strain ($\varepsilon$) & -2.0\% & -1.0\% & +1.0\% & +2.0\% \\
      \hline
      Space group & $Fmmm$ & $Fmmm$ & $Amam$ & $Amam$ \\
      Ni-planar O bond angle ($^{\circ}$)  & 0.9 & 0.1 & 7 & 9 \\
      Ni-apical O bond angle ($^{\circ}$)  & 0.0 & 0.0 & 14 & 16 \\
      In-plane axis (\AA) & 5.32 & 5.36 & 5.45 & 5.50 \\
      $c$ (\AA) & 20.73 & 20.50 &  20.18 & 20.00 \\
      Volume $a \times a \times c$\,(\AA$^3$) & 586.71 & 588.96 & 599.4 & 605\\
      
      \hline
    \end{tabular}
  \end{center}
\end{table}

\begin{figure*}[h]
    \includegraphics[clip=true,width=\columnwidth]{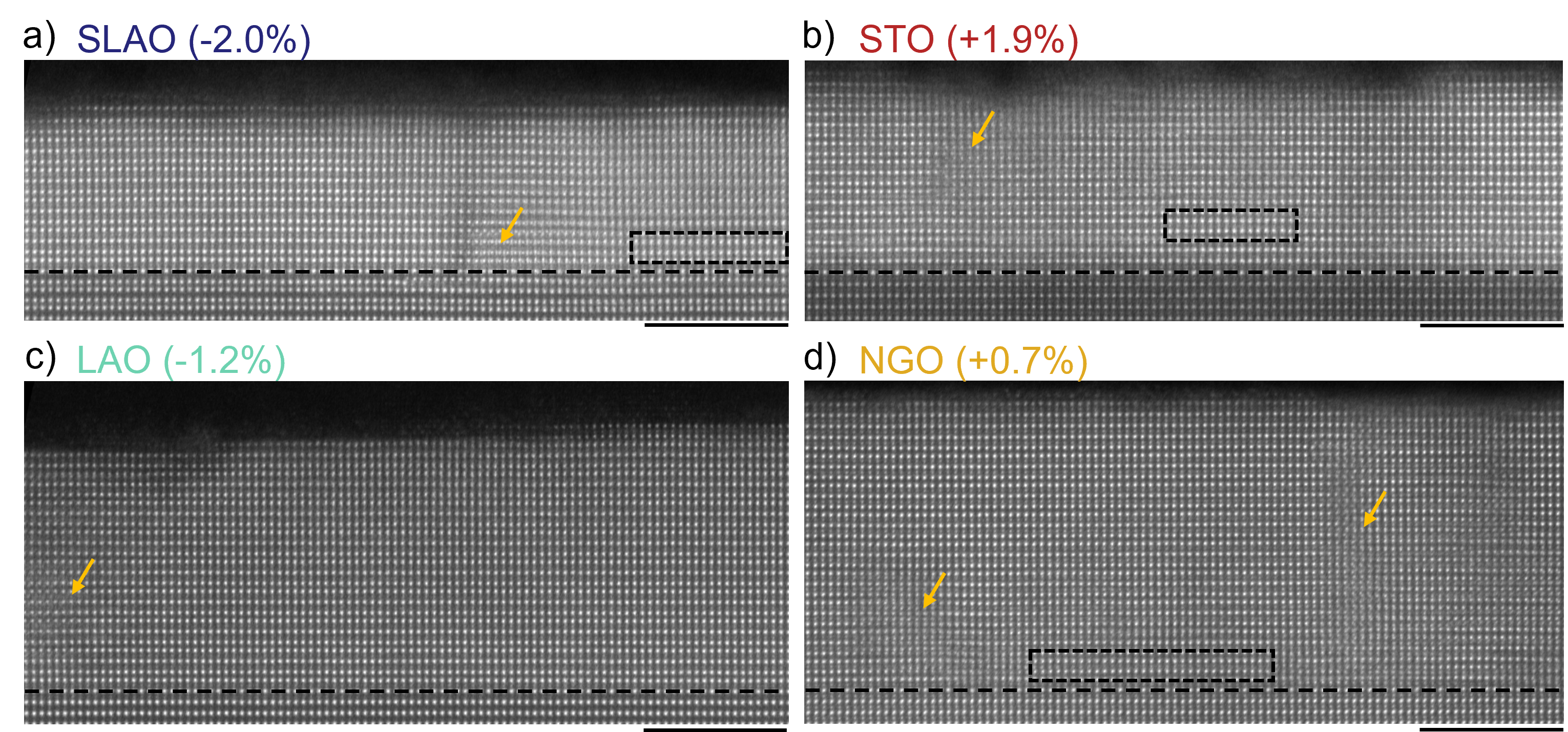}
    \caption{ \textbf{a-d)} Representative ADF-STEM images of \LNO thin films grown on SLAO (a), STO (b), LAO (c), and NGO (d). Black dashed lines mark the approximate position of the interface. Structural defects are indicated by yellow arrows and intergrowths of \LNOtwo by black boxes. Scale bar is 5 nm.
    }
     \label{fig:all_films}
\end{figure*}

\begin{figure*}[h]
    \includegraphics[clip=true,width=\columnwidth]{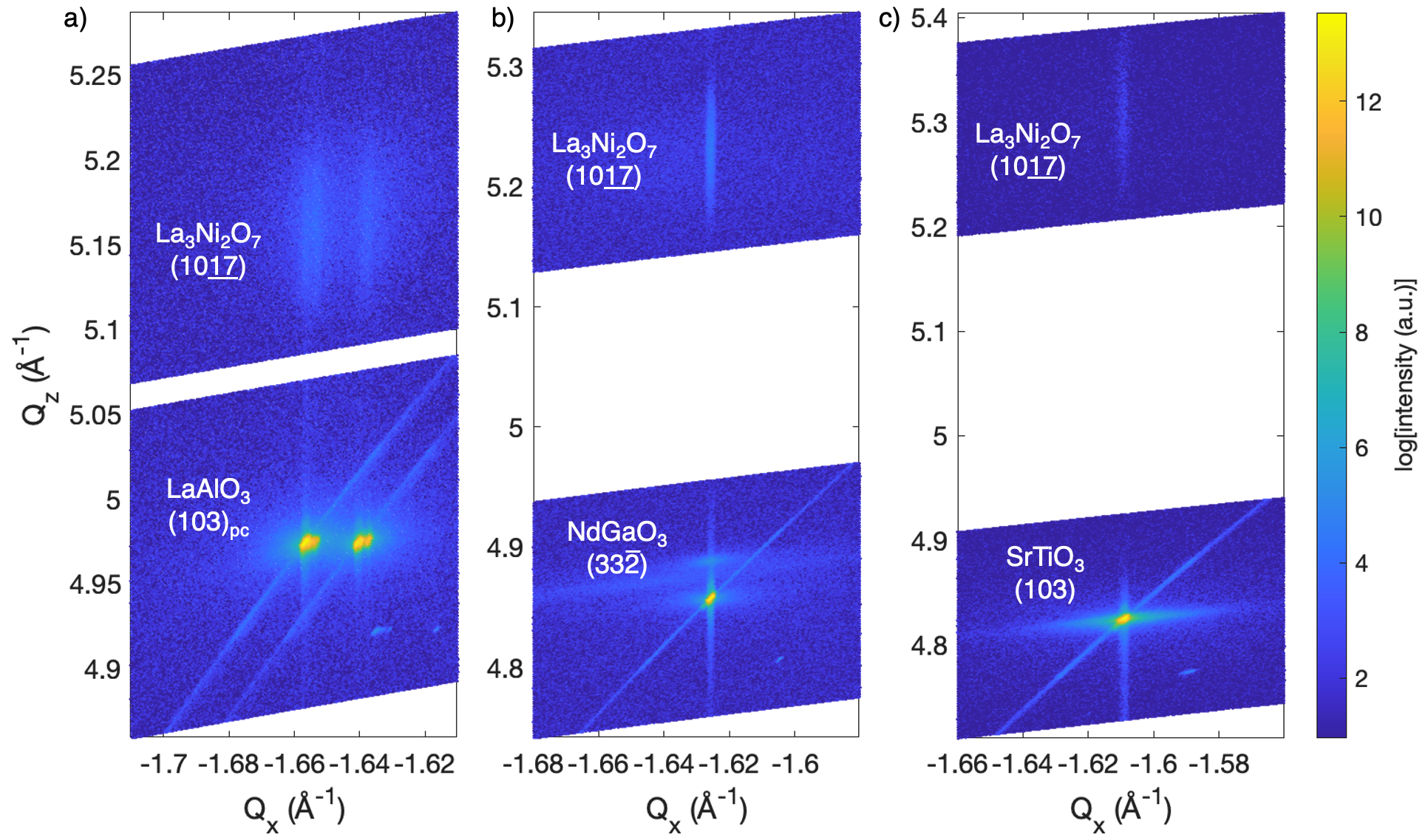}
    \caption{Reciprocal space maps (RSM) of \LNO thin films with the pseudo-tetragonal (10\underline{17}) Bragg peaks of \LNO alongside the following substrate peaks: \textbf{a)} (103)$_{pc}$ pseudo-cubic peak of LAO(100),  \textbf{b)} (33$\overline{2}$) peak of NGO(110), and  \textbf{c)} (103) peak of STO(001). In-plane lattice constants are extracted from the \LNO film peaks. RSM of \LNO on SLAO(001) is shown in extended data of \cite{ko2024signatures}. The LAO(100) sample exhibits two substrate peaks (and thus two signals from the strained \LNO) due to twinning in the substrate, and the left peak was used to extract the in-plane parameter based on the RSM measurement alignment.
    }
     \label{fig:RSMs}
\end{figure*}

\begin{figure*}[h]
    \includegraphics[clip=true, scale = 0.3]{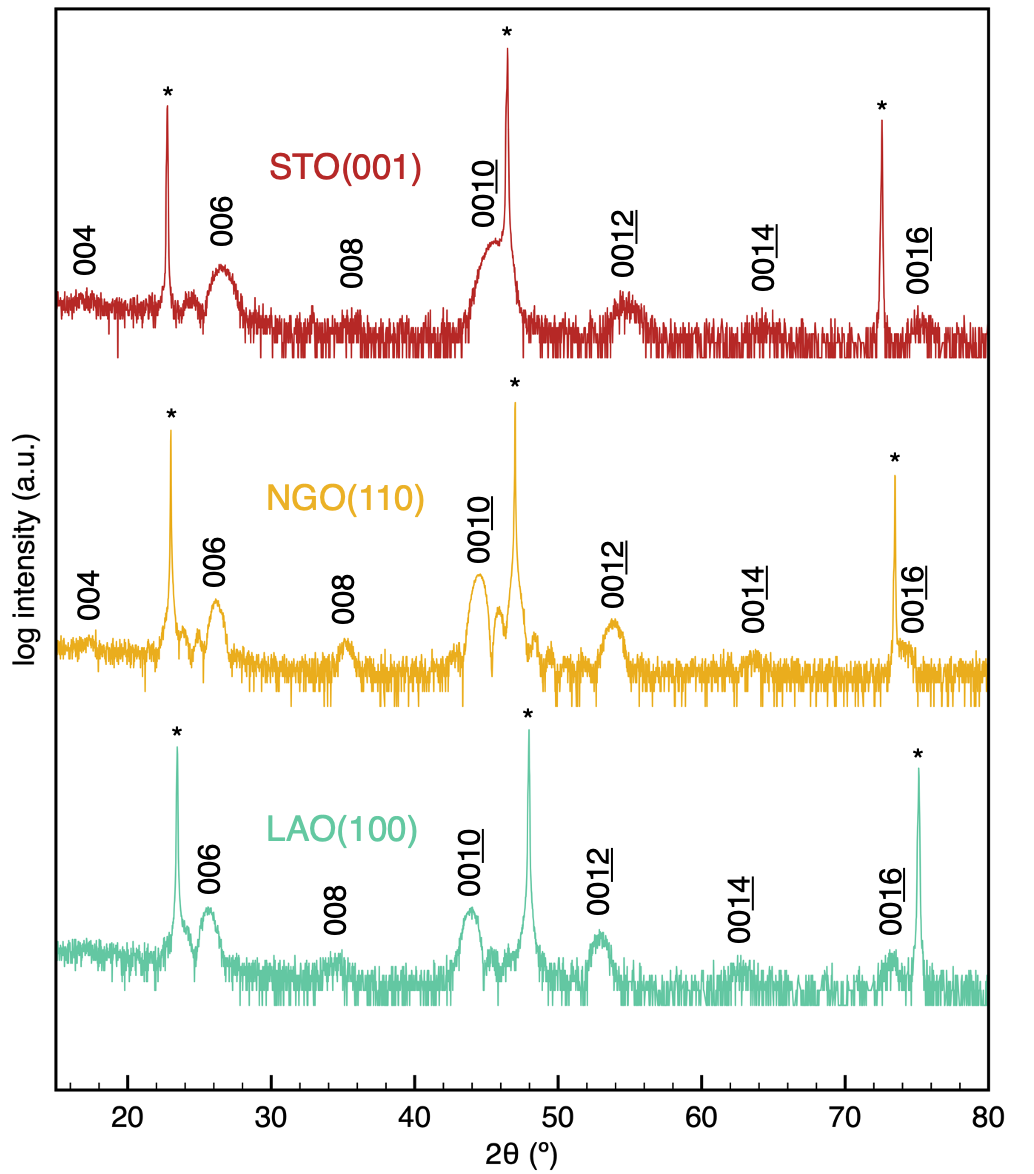}
    \caption{X-ray diffraction (XRD) of \LNO films on LAO(100), NGO(110), and STO(001) substrates. Substrate peaks are labeled with *. A Nelson-Riley fit of the indexed (00l) peaks of \LNO was used to extract the c-axis lattice parameters and corresponding error as measured by XRD \cite{nelson1945fit}. XRD of \LNO on SLAO(001) is shown in Figure 1 of \cite{ko2024signatures}.}  
     \label{fig:XRD}
\end{figure*}

\begin{figure*}[h]
    \includegraphics[clip=true, scale = 0.6]{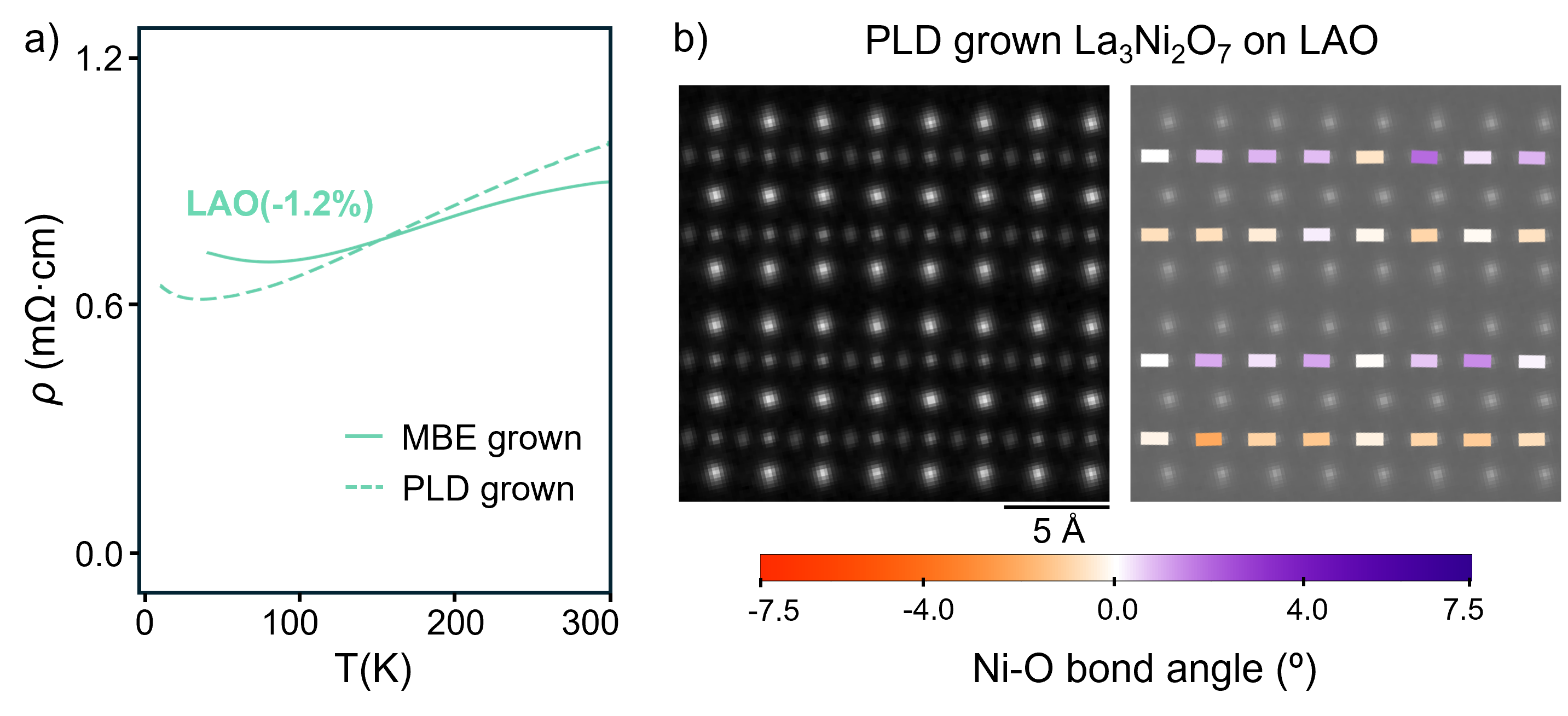}
    \caption{\textbf{a)} Resistivity $\rho$(T) of \LNO thin films epitaxially strained on LAO grown via molecular beam epitaxy (MBE) (solid line) and pulsed laser deposition (PLD) (dashed line) as described in Methods. The transport curves show similar behavior between the films grown using MBE and PLD. \textbf{b} Ptychographic reconstruction (left) and corresponding Ni-O bond angle map (right) of a PLD-grown \LNO film on LAO. The Ni-O coordination observed in MBE-grown \LNO films on LAO (Fig. 2b of the main text) are similar to the structure observed in the films grown by PLD. }
     \label{fig:PLDvsMBE_LAO}
\end{figure*}

\begin{figure*}[h]
    \includegraphics[clip=true,width=\columnwidth]{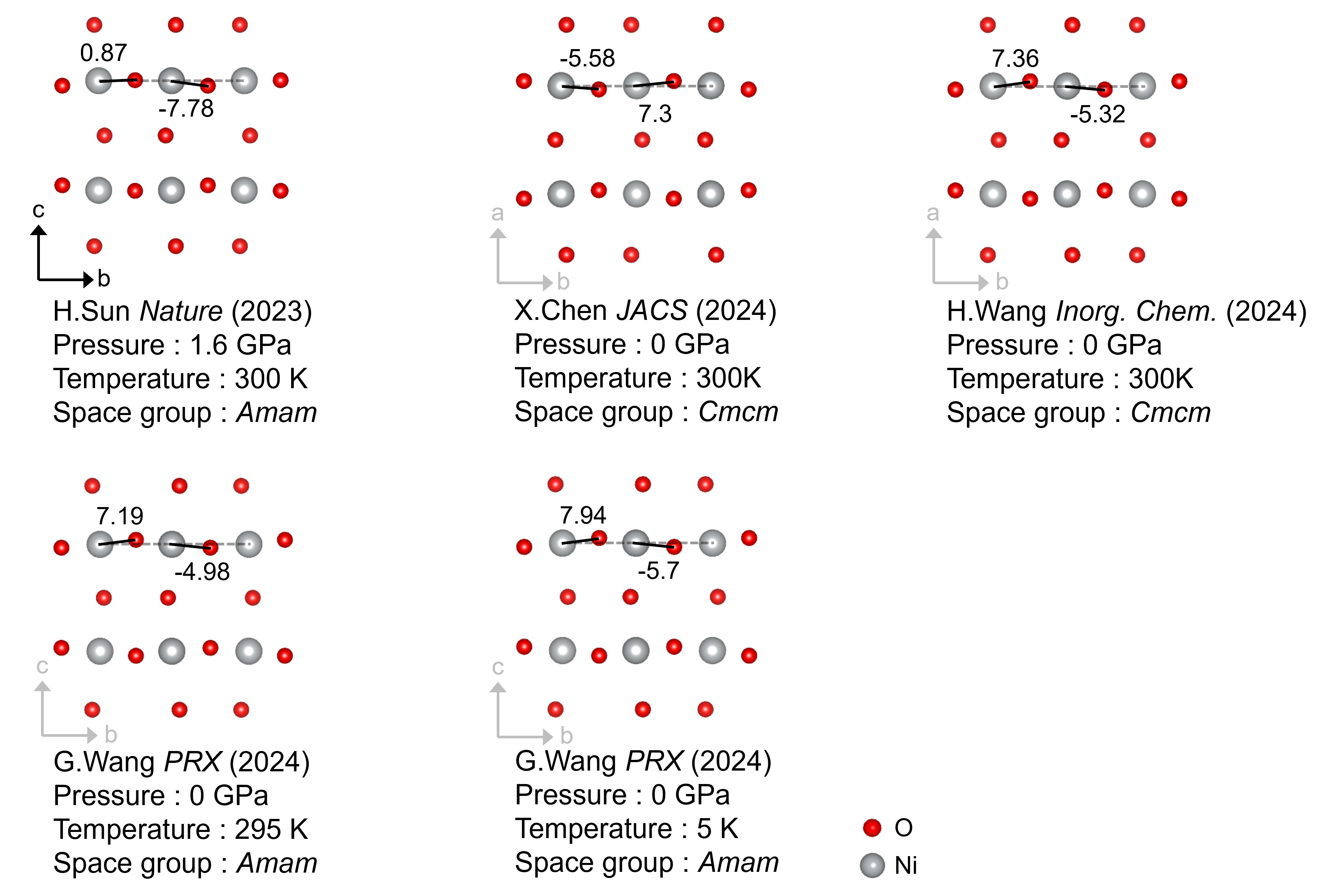}
    \caption{Various reported structures of \LNO at low pressure showing a single bilayer of the Ni-O$_6$ octahedra (La sites are omitted for clarity), showing variability in the assigned space groups and approximate measured Ni-planar O bond angles. Structures reported by \cite{sun2023signatures,chen2024polymorphism,HWangInorgChem2024,GWangPRX2024}.
    }
     \label{fig:LP structures}
\end{figure*}

\begin{figure*}[h]
    \includegraphics[clip=true,width=\columnwidth]{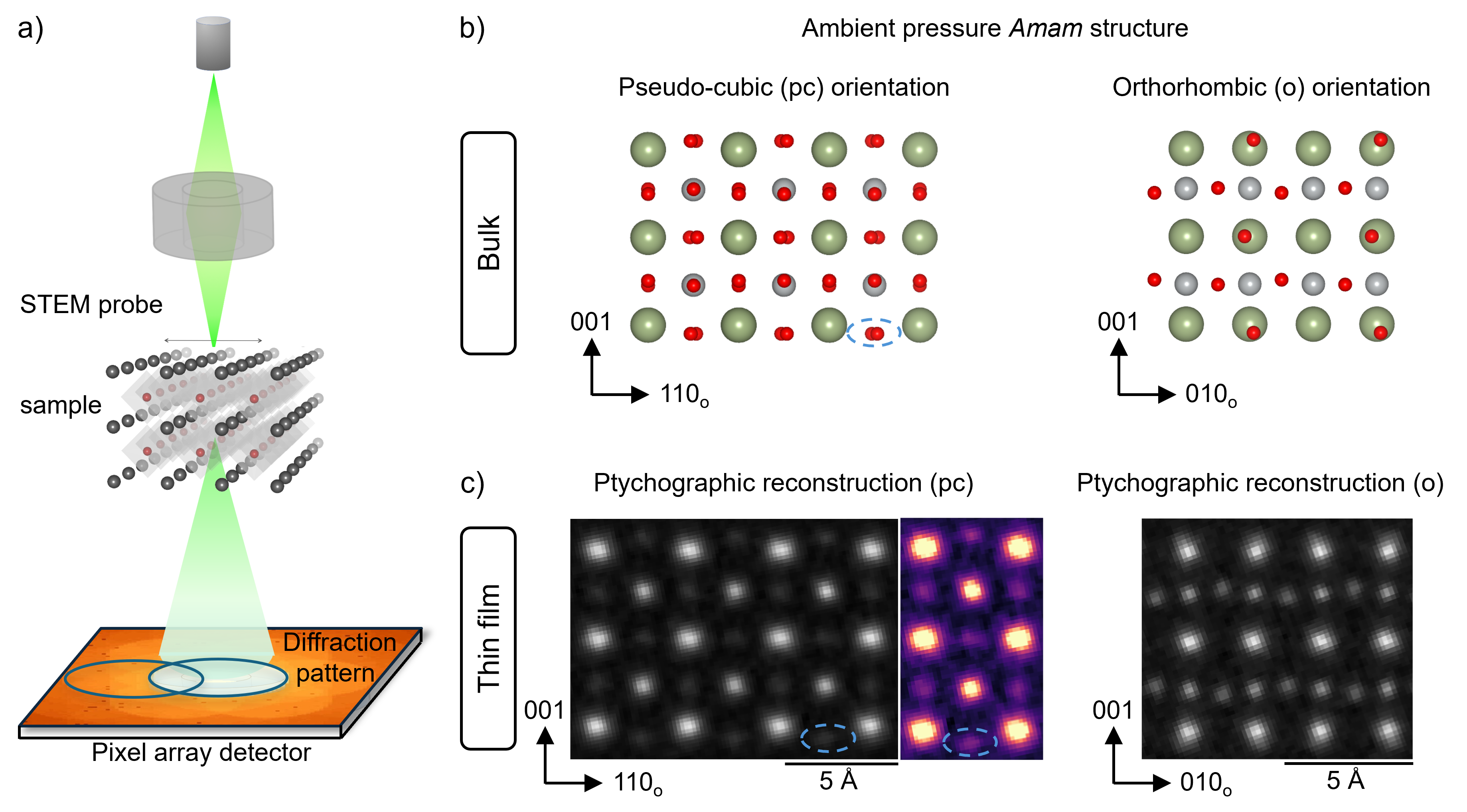}
    \caption{ \textbf{a)} Illustration of the experimental setup for multislice electron ptychography (MEP) showing a convergent STEM probe rastering across the sample. The scattered electron diffraction patterns are collected on a pixel array detector with high dynamic range \cite{EMPAD2}. \textbf{b)} Experimental bulk structure of a single bilayer as reported by \cite{sun2023signatures} along the pseudo-cubic [110]$_{o}$ (pc, left) and orthorhombic [100]$_{o}$ (o, right) projections. The oxygens are anti-phase stacked along the pseudo-cubic orientation, as highlighted by the blue circle \cite{zhen327Nature}. \textbf{c)} Experimental multislice electron ptychographic reconstruction of an \LNO thin film on STO along both pseudo-cubic (left) and orthorhombic (right) projections. A slight elongation of the oxygen columns due to the anti-phase stacking is visible along the pc orientation. The antiphase stacking also leads to a reduction in the intensity of oxygen columns compared to La and Ni columns. A saturated and cropped magma-colored field-of-view of the reconstruction is given for clear visibility of oxygen columns along the pc projection. 
    }
     \label{fig:Ptycho}
\end{figure*}

\begin{figure*}[h]
    \includegraphics[clip=true,scale = 0.7]{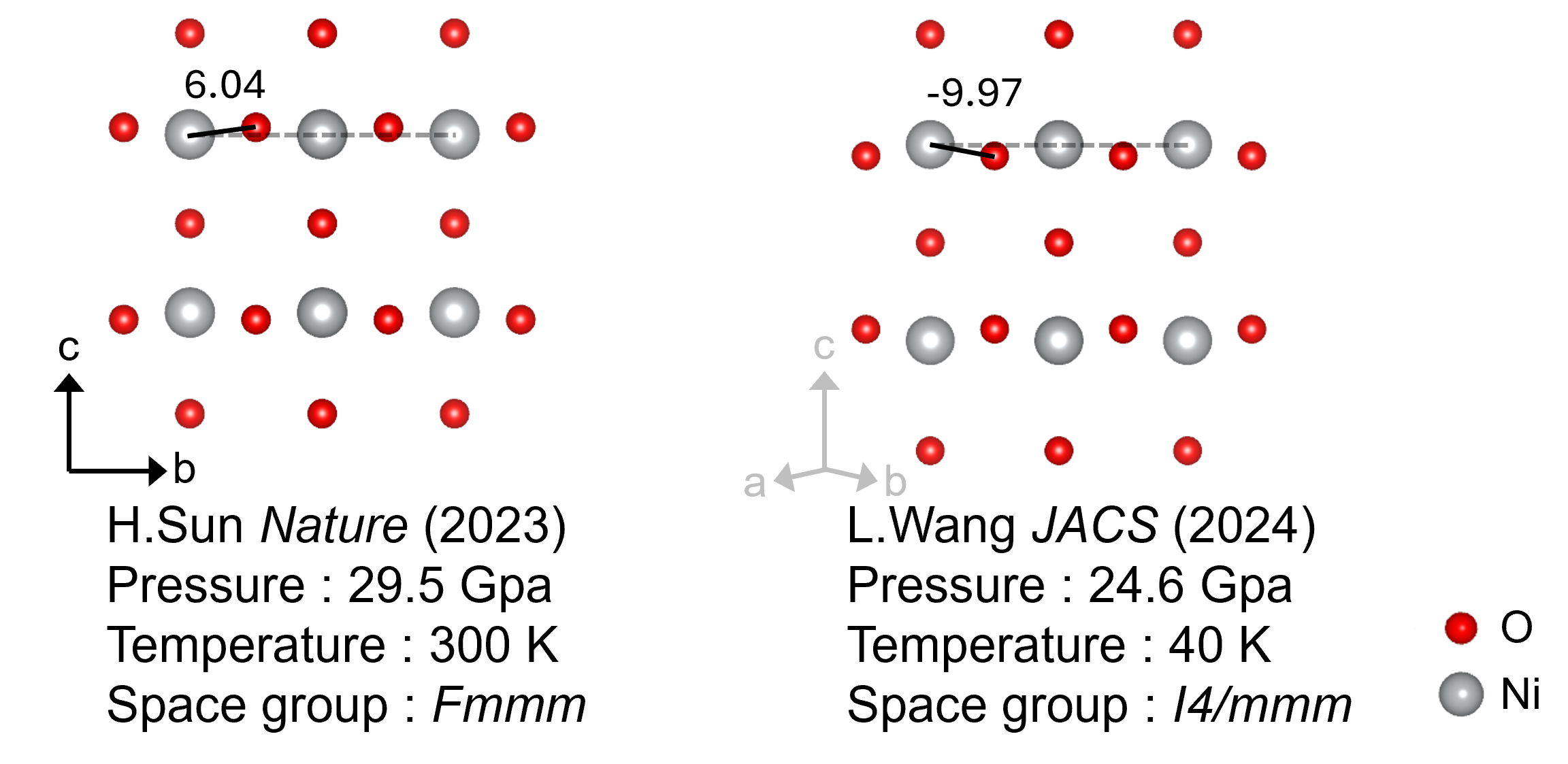}
    \caption{Two reported structures of \LNO at high-pressure and different temperatures, here showing only a single bilayer of the Ni-O$_6$ octahedra (La sites are omitted for clarity). Approximate Ni-planar O bond angles are noted. Beyond their space group assignments ($Fmmm$ and $I4/mmm$), the structures also differ by the distortion direction of the planar O either away from (left) or towards (right) the center plane of the bilayer. Structures reported by from \cite{sun2023signatures, wang2024structure}.
    }
     \label{fig:HP structures}
\end{figure*}

\begin{figure*}[h]
    \includegraphics[clip=true,width=\columnwidth]{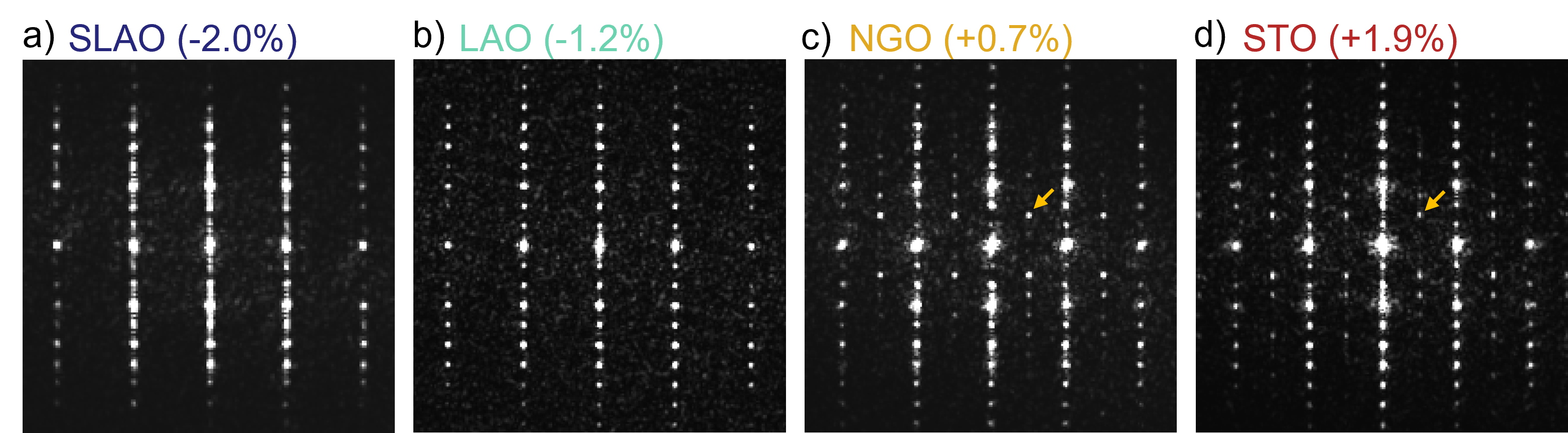}
    \caption{\textbf{a-d)} Fourier transforms (FTs) of ADF-STEM images of \LNO films grown on SLAO (a), LAO (b), NGO (c), and  STO (d). Extra peaks (yellow arrow) are visible in the tensile strained films and absent in the compressive films indicating structural symmetries consistent with $Amam$ and $Fmmm$ or $I4/mmm$, respectively.
    }
     \label{fig:FFT}
\end{figure*}

\begin{figure*}[h]
    \includegraphics[clip=true,scale=.8]{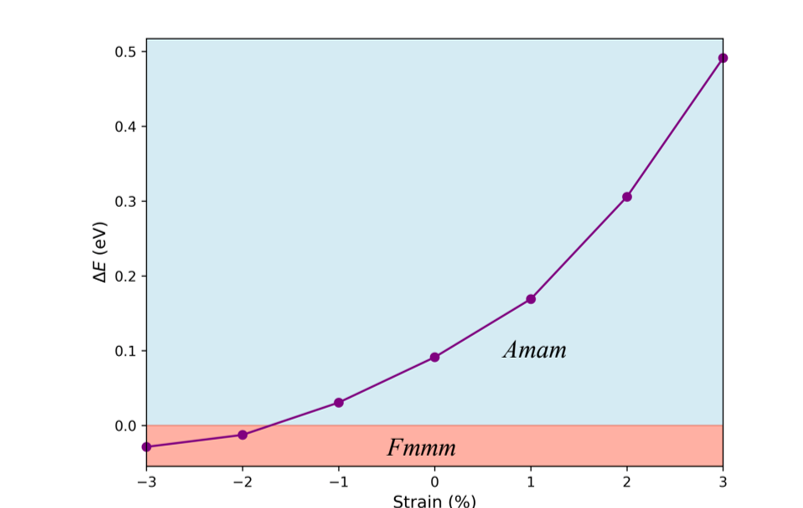}
    \caption{ Calculated energies of \LNO under biaxial strains ranging from -3\% to +3\%, showing a strain-dependent crossover of the lowest-energy crystal symmetry from $Amam$ under tensile and zero strain to $Fmmm$ under compressive strain. Here we emphasize the clear trend towards higher symmetry $Fmmm$ with increasing compressive strain as was similarly reported by \cite{rhodes2024structural} rather than the quantitative values of the calculated strain at the phase boundary which can be sensitive to computational parameters of the calculations.}
     \label{fig:DFT}
\end{figure*}

\begin{figure*}[h]
    \includegraphics[clip=true, scale = 0.6]{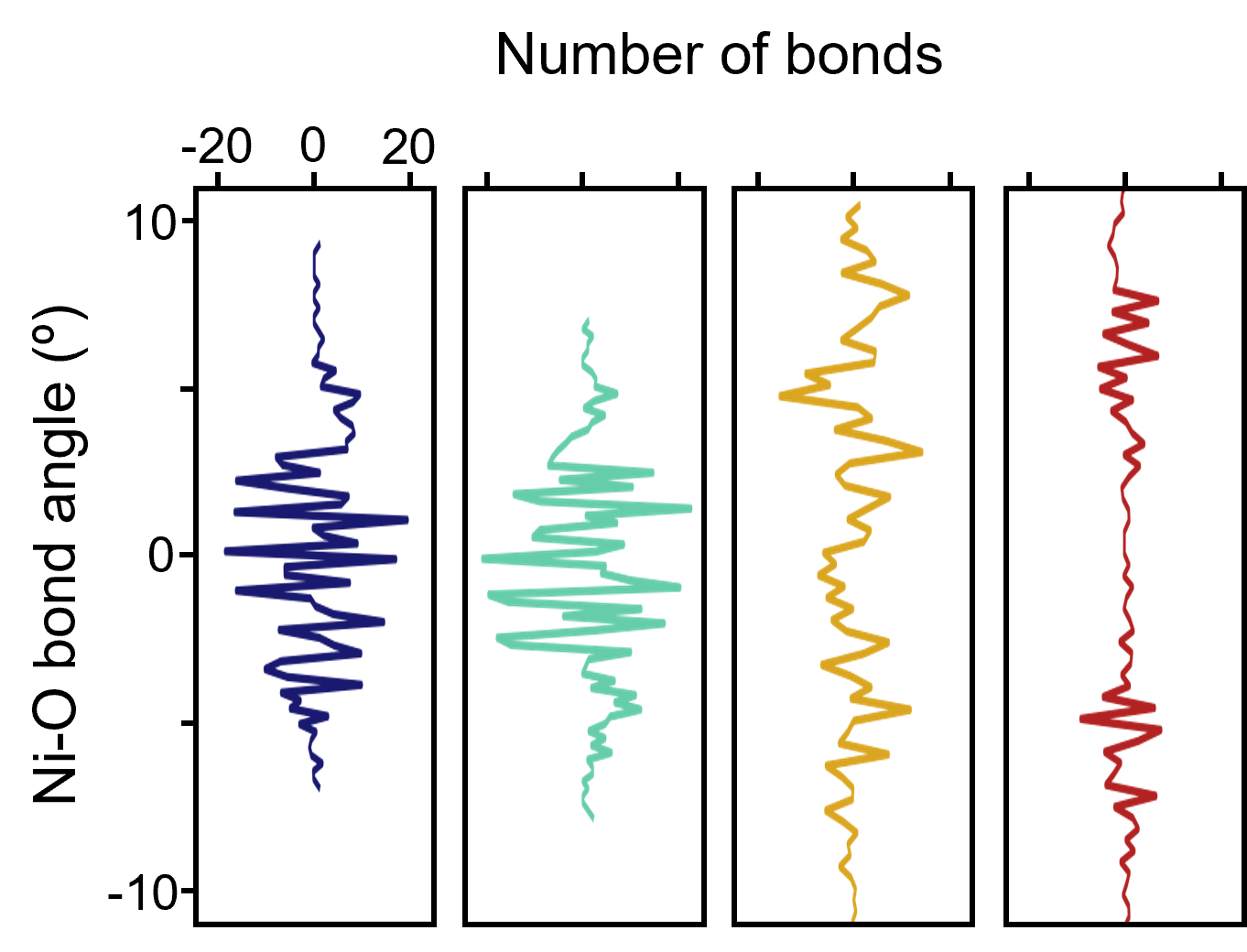}
    \caption{Residuals to the two-Gaussian fits of Ni-O bond angle histograms in Fig. 2f of the main text. }
     \label{fig:ResFit}
\end{figure*}

\begin{figure*}[h]
    \includegraphics[clip=true,width=\columnwidth]{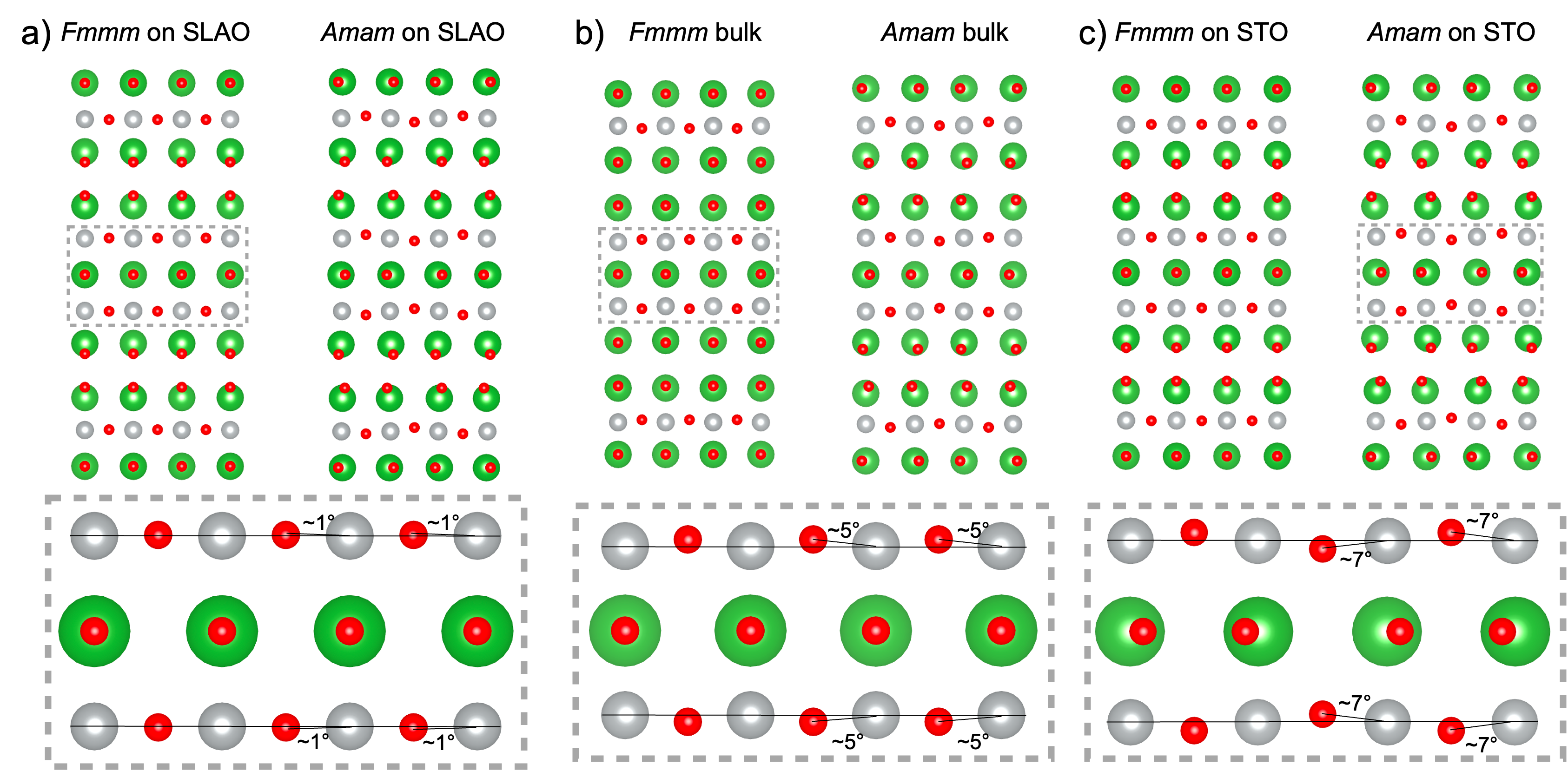}
    \caption{Structural relaxations of reported $Fmmm$ and $Amam$ bulk \LNO calculated by density functional theory (DFT) with Hubbard $U$. Starting from the experimental bulk structures reported by Sun, et al. \cite{sun2023signatures} (b), in-plane lattice constants are constrained to match SLAO (a) and STO (c) substrates. Close inspection of the two Ni-O planes within a single bilayer (dashed boxes and enlarged insets) can be used to estimate the Ni-planar bond angles for comparison to experimental MEP. Note that because the $Fmmm$ structure is measured at 29.5 GPa, the compressed in-plane lattice constants are smaller than those constrained to the SLAO substrate such that the structure ``expands'' slightly on SLAO, which may explain the shallower Ni-planar O bond magnitudes in the thin film compared to the pressurized bulk. Also, note that a simple calculation of compressive strain starting from the $Amam$ structure does not converge to the $Fmmm$-like octahedral symmetry which is observed in thin films measured by MEP, but rather retains the $Amam$-like rotation pattern. 
    }
     \label{fig:edgar-relax}
\end{figure*}

\begin{figure*}[h]
    \includegraphics[clip=true, scale = 0.8]{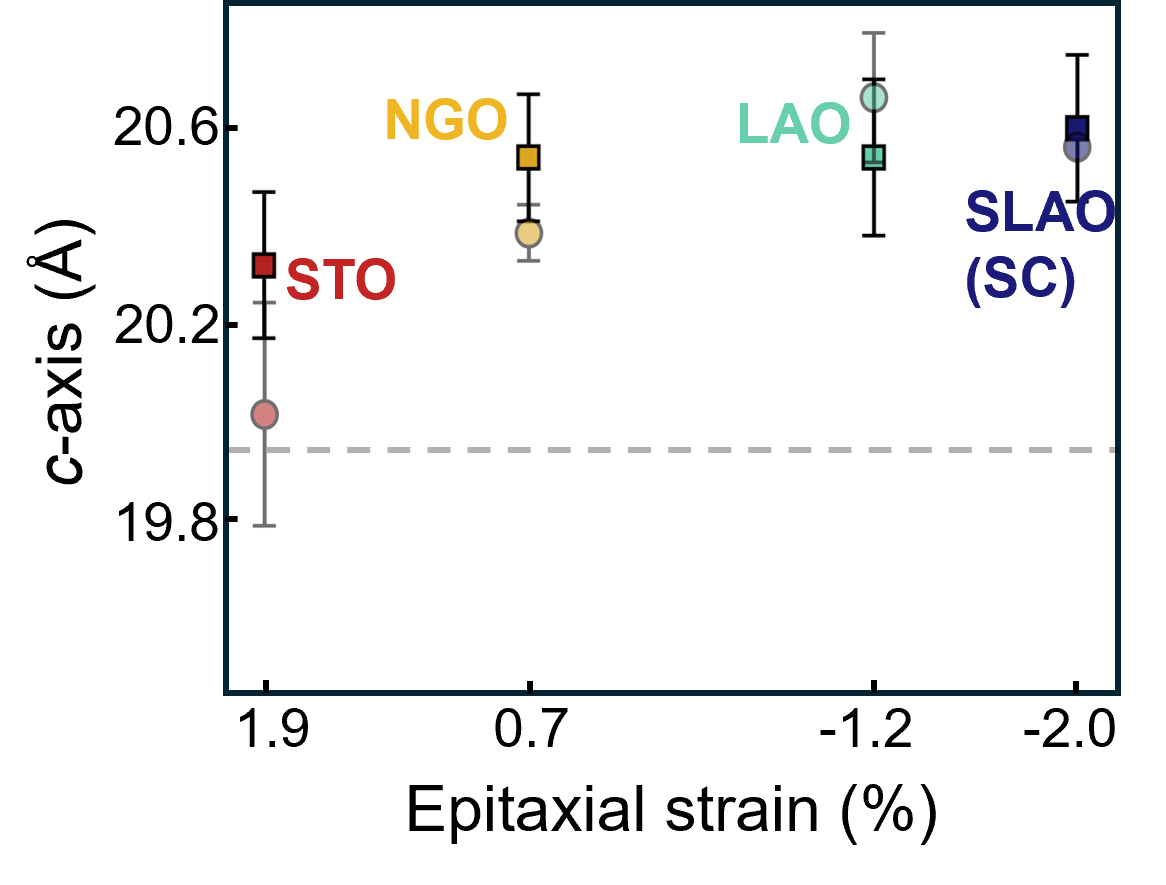}
    \caption{$c$-axis lattice constants of \LNO thin films as a function of nominal epitaxial strain measured via XRD (circles) and ADF-STEM (squares). Errors from Nelson-Riley fits of the XRD data for films on STO, NGO, and LAO are plotted along with the standard deviation of the ADF-STEM measurements. XRD and STEM measurements of $c$-axis lattice spacing agree within the uncertainty of the measurements.}
     \label{fig:xrderr}
\end{figure*}

\begin{figure*}[h]
    \includegraphics[clip=true,width=\columnwidth]{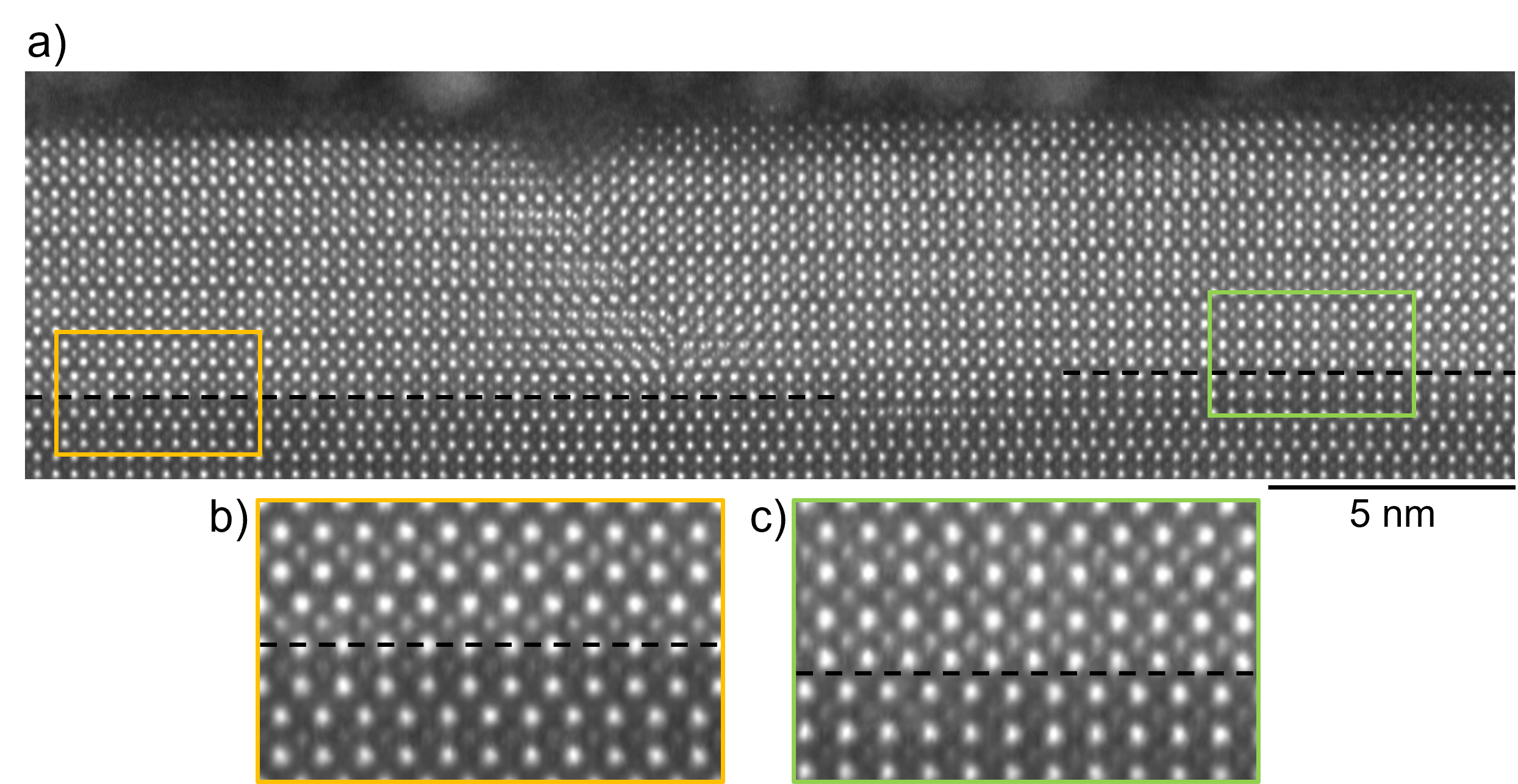}
    \caption{ \textbf{a)} ADF-STEM image of \LNO film grown on SLAO along the pseudo-cubic [110]$_{o}$ projection. Black dashed line indicates the approximate position of interface inferred from changes in the intensity of atomic columns. The interfacial structure varies from the left side of the image to the right with a step edge in the middle. The two distinct interfacial structures are highlighted with yellow and green boxes. \textbf{b,c)} Enlarged view of the interfaces in yellow (b) and green (c) boxes. In (b), the interface between substrate and film appears in the middle of the bilayer followed by a single layer of \LNOtwo while in (c) there is a Ruddlesden-Popper gap at the interface and the first layer of the film forms as trilayer \LNOfour.}
     \label{fig:interface}
\end{figure*}

\begin{figure*}[h]
    \includegraphics[clip=true,scale=.8]{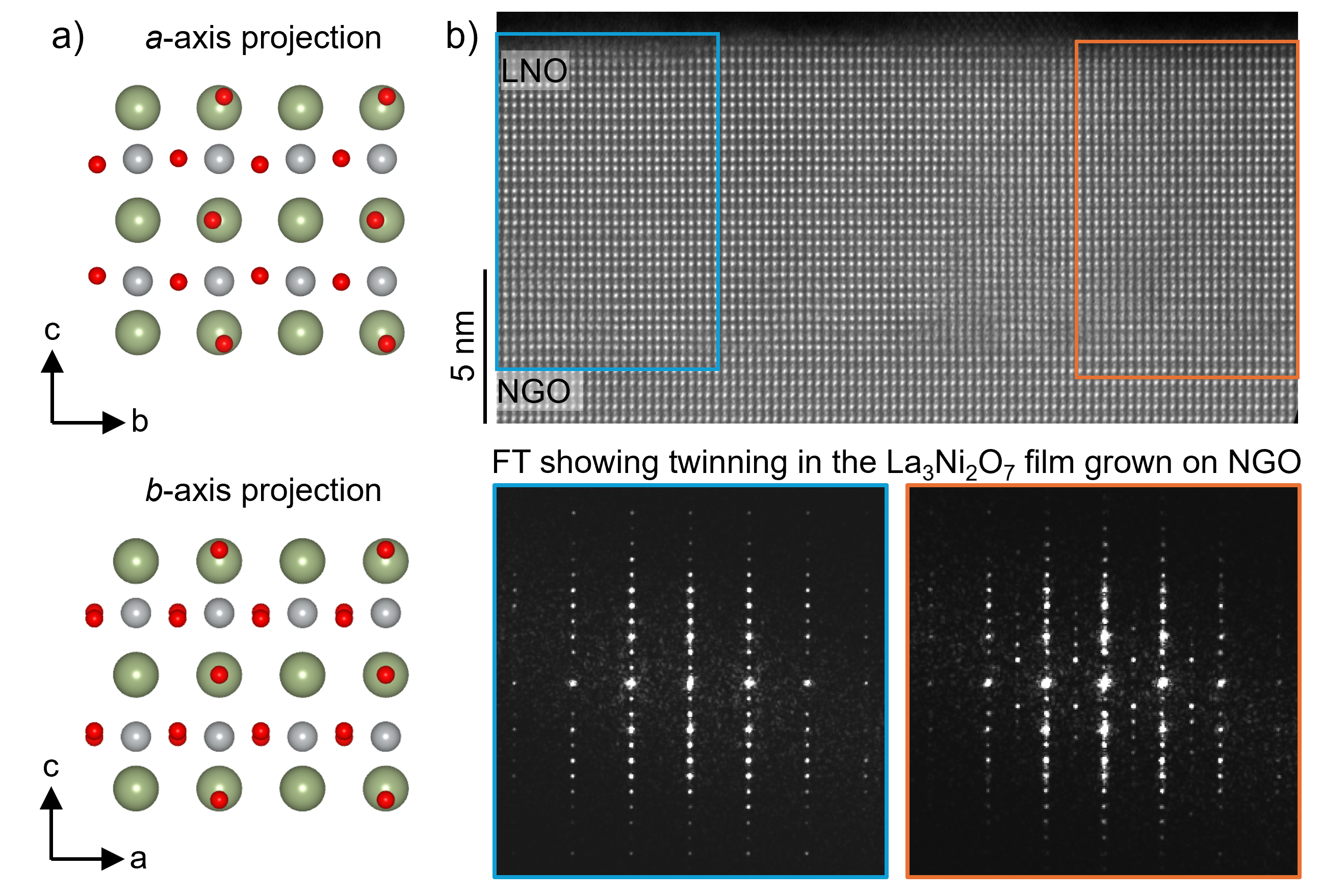}
    \caption{ \textbf{a)} Experimental bulk structure of a single bilayer as reported by \cite{sun2023signatures} along the a-axis projection (top) and b-axis projection (bottom). \textbf{b)} ADF-STEM image (top) of \LNO grown on NGO. Fourier transforms (FTs) of two different areas of the film marked by blue and orange boxes show distinct peaks. The orange region has half-order peaks that are lacking in the blue region which are consistent with the inequivalent $a$ (orange) and $b$ (blue) projections of the $Amam$-like structure, indicating the presence of twinning in the film. Measurements of Ni-planar O bond angles by MEP in the main text is limited to regions with $a$-axis projection.
    }
     \label{fig:NGOTwinning}
\end{figure*}

\begin{figure*}[h]
    \includegraphics[clip=true, scale = 0.6]{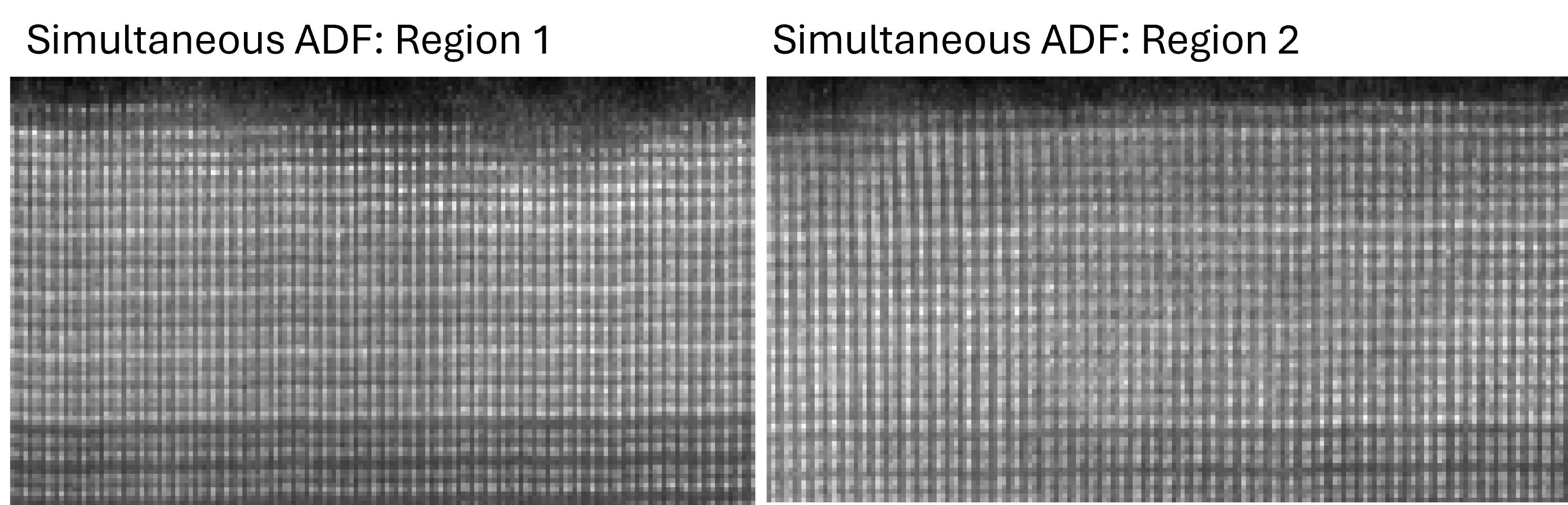}
    \caption{ADF images of Region 1 and Region 2 acquired simultaneously with the EEL spectra shown in Fig. 4d of the main text. Both ADF images confirm the overall crystalline quality and majority adherence to \LNO stacking. Note that the reduced signal-to-noise ratio of the images is due to the low probe currents used for EELS measurements and apparent ``distortions'' to the lattice are from small amounts of sample drift during the long acquisition times. }
     \label{fig:SimADF_EELS}
\end{figure*}

\end{document}